\documentclass[sigconf]{acmart}

\usepackage{amsmath,amsfonts}
\usepackage{lettrine}
\usepackage{subfigure}
\usepackage{multirow}
\usepackage{multicol}
\usepackage{tablefootnote}
\usepackage{enumerate}
\usepackage{enumitem}
\usepackage{extarrows}
\usepackage{booktabs}
\usepackage{mathtools}
\usepackage{multirow}
\usepackage{caption}
\usepackage{graphicx}
\usepackage{url} 
\usepackage{pifont}
\usepackage{extarrows}
\usepackage{amsfonts}
\usepackage{marvosym}

\usepackage[ruled,longend, algo2e]{algorithm2e}
\usepackage{algorithm}
\usepackage{algorithmic}

\usepackage{listings}
\usepackage{xcolor}
\AtBeginDocument{%
  }

\setcopyright{acmlicensed}
\copyrightyear{2018}
\acmYear{2018}
\acmDOI{XXXXXXX.XXXXXXX}
\acmConference[Conference acronym 'XX]{Make sure to enter the correct
  conference title from your rights confirmation email}{June 03--05,
  2018}{Woodstock, NY}
\acmISBN{978-1-4503-XXXX-X/2018/06}

\begin{document}

\title{Learning Multi-Branch Cooperation for Enhanced Click-Through Rate Prediction at Taobao}

\author{Xu Chen, Zida Cheng, Yuangang Pan, Shuai Xiao, \\ Xiaoming Liu, Jinsong Lan, Xiaoyong Zhu, Bo Zheng, Ivor W. Tsang}
\email{{huaisong.cx,chengzida.czd,shuai.xsh,jinsonglan.ljs,xiaoyong.z,bozheng}@alibaba-inc.com, xm.liu@xjtu.edu.cn,{pan_yuangang,ivor_tsang}@cfar.a-star.edu.sg}
\affiliation{%
  \institution{Alibaba Group, Xi'an Jiao Tong University, China and A$^*$ STAR, Singapore}
  \country{}
}

\renewcommand{\shortauthors}{Xu Chen et al.}

\begin{abstract}
  Existing click-through rate (CTR) prediction works have studied the role of feature interaction through a variety of techniques. Each interaction technique exhibits its own strength, and solely using one type usually constrains the model's capability to capture the complex feature relationships, especially for industrial data with enormous input feature fields. Recent research shows that effective CTR models often combine an MLP network with a dedicated feature interaction network in a two-parallel structure. However, the interplay and cooperative dynamics between different streams or branches remain under-researched. In this work, we introduce a novel Multi-Branch Cooperation Network (MBCnet) which enables multiple branch networks to collaborate with each other for better complex feature interaction modeling. Specifically, MBCnet consists of three branches: the Extensible Feature Grouping and Crossing (EFGC) branch that promotes the model's memorization ability of specific feature fields, the low rank Cross Net branch and Deep branch to enhance explicit and implicit feature crossing for improved generalization. Among these branches, a novel cooperation scheme is proposed based on two principles: \textbf{\textit{Branch co-teaching}} and \textbf{\textit{moderate differentiation}}. \textbf{\textit{Branch co-teaching}} encourages well-learned branches to support poorly-learned ones on specific training samples. \textbf{\textit{Moderate differentiation}} advocates branches to maintain a reasonable level of difference in their feature representations on the same inputs. This cooperation strategy improves learning through mutual knowledge sharing via co-teaching and boosts the discovery of diverse feature interactions across branches. Extensive experiments on large-scale industrial datasets and online A/B test at Taobao app demonstrate MBCnet's superior performance, delivering a 0.09 point increase in CTR, 1.49\% growth in deals, and 1.62\% rise in GMV. Core codes are available online.
\end{abstract}

\ccsdesc[500]{Information systems~Retrieval models and ranking}

\keywords{multi-branch cooperation, CTR prediction, feature grouping, branch co-teaching, moderate differentiation}


\maketitle

\begin{figure}[t]
\centering
\begin{minipage}[t]{0.24\textwidth}
\centering
\includegraphics[width=\textwidth]{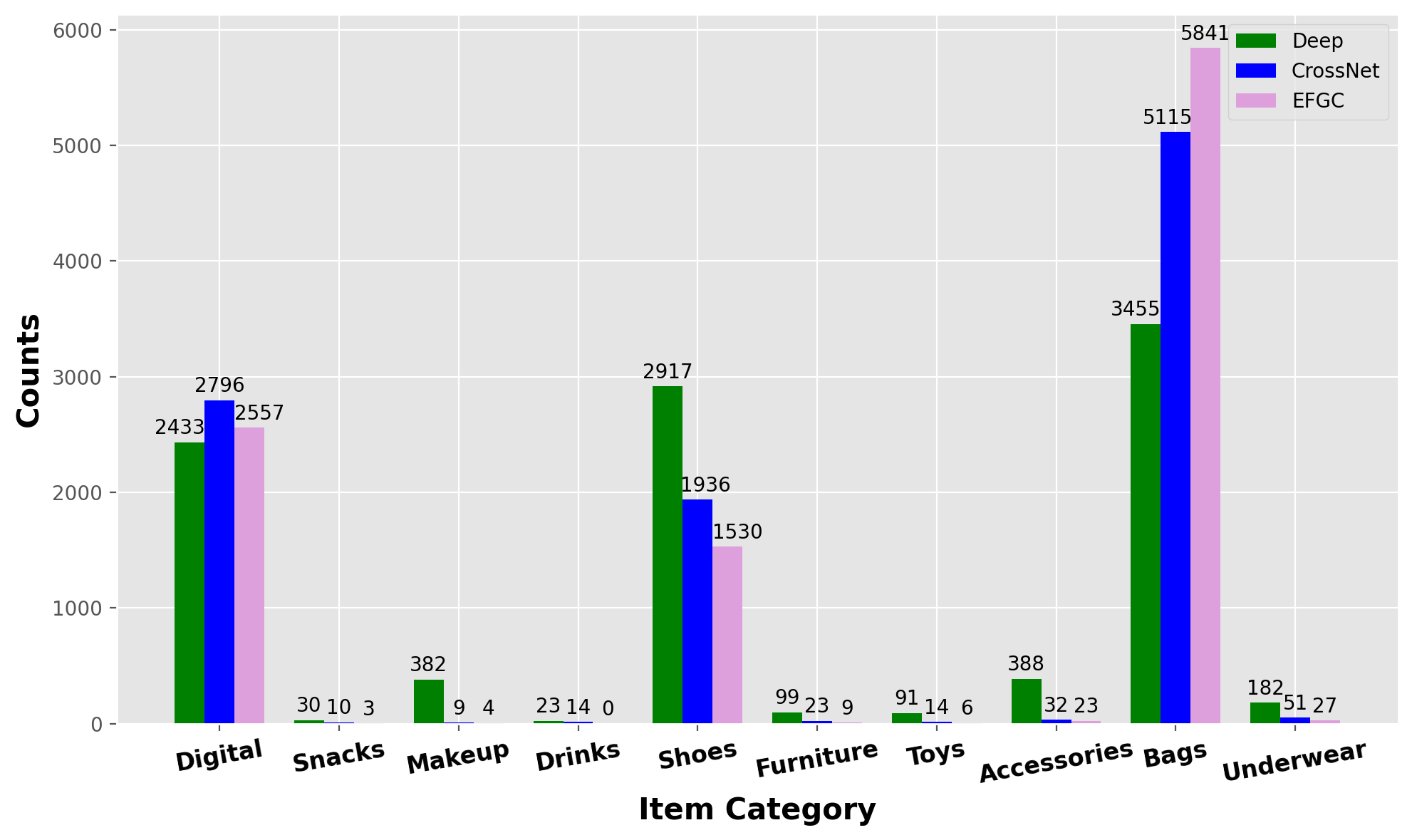}
\vspace{-20pt}
\caption*{\small{(a) Category Distribution}}
\end{minipage}
\hspace{5pt}
\begin{minipage}[t]{0.21\textwidth}
\centering
\includegraphics[width=\textwidth]{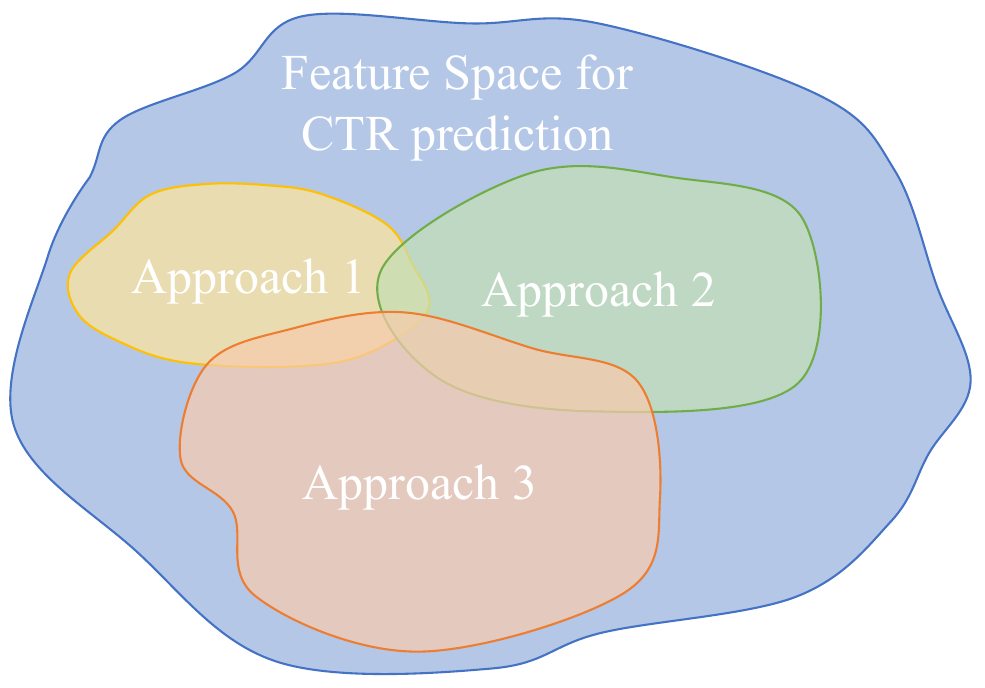}
\vspace{-20pt}
\caption*{\small{(b) Feature Space Coverage}}
\end{minipage}
\vspace{-12pt}
\caption{(a) illustrates the category distribution of top-10k most accurately learned samples (\textit{i.e.}, those with low logloss) across different feature interaction techniques in our MBCnet. (b) gives an example to illustrate the feature space for CTR prediction and the coverage of different approaches.}
\label{figure:taobao_data_example}
\vspace{-8pt}
\end{figure}
\section{Introduction}
Click-Through Rate (CTR) prediction which estimates the probability of a user clicking on a candidate item, is a fundamental task in online services like recommendation, retrieval, and advertising~\cite{cheng2016wide,zhou2018deep,ouyang2019deep}.
The precision in predicting CTR not only substantially impacts user engagement but also has a significant effect on the revenue of industrial businesses. One key aspect of developing precise CTR prediction is to capture complex feature interactions, which enables the model to generalize across various scenarios.

For many years, researchers have examined the role of feature interaction in CTR by employing a range of methodologies. Earliest algorithm is Logistic Regression (LR)~\cite{7154880}, which is a linear model and heavily relies on hand-crafted input features by domain experts. Factorization Machine (FM)~\cite{5694074} combines polynomial regression models with factorization techniques to automatically capture effective feature interactions. These methods primarily focus on low-order feature interactions and have shown limitations in capturing complex patterns among features~\cite{YANG2022102853}. Due to the superior non-linear feature learning ability of Deep Neural Networks (DNN), many scholars have investigated deep learning techniques on CTR prediction~\cite{9277337}. For example, Wide \& Deep~\cite{cheng2016wide} consists of a jointly trained wide linear model and deep neural networks, combining the benefits of memorization and generalization for recommender systems. DeepFM~\cite{10.5555/3172077.3172127} and xDeepFM~\cite{10.1145/3219819.3220023} combine the power of factorization machines and deep learning, to emphasize both low- and high-order feature interactions. 
Instead of employing FM to model feature relationships, \citet{wang2021dcn,10.1145/3124749.3124754} introduced a novel cross network that explicitly performs feature crossing with each layer, which eliminates the need for manual feature engineering.
Further, \citet{Mao023} observed that even two MLP networks in parallel can achieve surprisingly good performance compared to many well-designed models. Thereby, they developed a simple yet strong dual MLP model to capture diverse feature interactions and this model has achieved state-of-the-art performance on public benchmarks.

\textit{Each feature interaction technique brings its own advantages, and solely depending on one may hinder the model's potential to capture complex feature relationships.} Particularly in industrial contexts with vast numbers of users and items, the data patterns can be exceedingly intricate. We usually incorporate hundreds of input feature fields to learn for the data patterns, but such many feature fields also exhibit enormous useful feature interactions. The modeling ability of single feature interaction technique is limited and insufficiently effective~\cite{10.1145/3041021.3054192}. We analyze the difference of different feature interaction techniques on our industrial data.
Figure~\ref{figure:taobao_data_example} (a) shows the category distribution of the top-10k most accurately predicted samples (\textit{i.e.}, those with low logloss) across different feature interaction techniques in MBCnet. Figure~\ref{figure:taobao_data_example} (b) illustrates the feature space for CTR prediction and the coverage of different branches. 
Through these figures, we see different feature interaction approaches have different modeling abilities. Relying on one single approach cannot well cover the complex high-order feature space for CTR prediction.
Recent researchers have highlighted that existing successful CTR models~\cite{cheng2016wide,wang2021dcn,10.1145/3124749.3124754,10.5555/3172077.3172127,10.1145/3219819.3220023,10.1145/3357384.3357925} usually adopt a two-branch architecture and work in an ensemble style. The ensemble style permits the model to learn feature interactions from different perspectives. 
\textit{Despite the success of above models, the interplay and cooperative dynamics between different streams or branches remain under-researched.} Recent models ensemble different techniques for CTR prediction by prediction voting or latent feature mean pooling or concatenation, which means no explicit interplay and cooperative signals are provided. This could largely limit the whole model's learning ability, especially when majority branches have wrong predictions. It motivates us to study the working principles of multi-branch CTR networks and develop a more effective model to capture the complex patterns.

In this work, we propose a novel Multi-Branch Cooperation Network (MBCnet) which enables multiple branch networks to collaborate with each other for better complex feature interaction modeling. MBCnet consists of three different prediction branches and a cooperation scheme. One important prediction branch is a designed Extensible Feature Grouping and Crossing (EFGC) module, which groups hundreds of feature fields and only conducts feature interaction between specific group pairs by domain-driven knowledge. It is extensible to design any intended feature field interactions, and also efficient to avoid redundant feature interactions. This branch enhances the model's memorization ability to capture domain-driven feature interactions for CTR prediction. The other two branches are the existing popular and powerful low-rank Cross Net~\cite{wang2021dcn} for explicit feature crossing and Deep Net~\cite{10.1145/3124749.3124754,10.5555/3172077.3172127} for implicit feature crossing, which both improve the model's generalization ability~\cite{cheng2016wide}. These three branches exhibit diverse characteristics and modeling capabilities, which can be integrated into a more powerful model. 
More importantly, we propose a multi-branch cooperation scheme based on two principles to better exert advantages of different branches. \emph{The first principle (\textbf{\textit{branch co-teaching}}) dictates that well-learned branch should assist poorly-learned branch on particular samples, while the second principle (\textbf{\textit{moderate differentiation}}) maintains that latent features of distinct branches should preserve a moderate level of differentiation on the same input samples, avoiding extremes of either excessive divergence or excessive similarity}. Specifically, the first principle is embodied in a co-teaching objective function, where the well-learned branch uses its predictions as pseudo labels to guide the poorly-learned one on disagreed predicted samples. The second principle is formulated as an equivalent transformation regularization loss between latent features of branches. 
\emph{The cooperative strategy promotes learning abilities of different branches through knowledge transfer via co-teaching and also enhances their capabilities to uncover various patterns of feature interactions}. 

Through experiments, we have shown that MBCnet can largely improve CTR prediction performance on our industrial datasets. We have conducted an online A/B test in image2product search\footnote{It is Alibaba's online service that provides product search via image uploads.} at Taobao app\footnote{https://www.taobao.com/}, which revealed that MBCnet could achieve an absolute \textbf{0.09 point} CTR improvement, along with a relative \textbf{1.49\%} deal growth and a relative \textbf{1.62\%} Gross Merchandise Value (GMV) rise. The contributions are summarized as follows:
\begin{itemize}
    \item We propose a novel multi-branch cooperation network (MBCnet) which ensembles different feature interaction branches to enlarge the capacity of capturing complex feature patterns in industrial data. Specifically, MBCnet consists of a designed EFGC branch and two popular and powerful branches. Each of them possess unique capabilities and strengths, enabling them to form a more powerful CTR prediction model.
    \item We introduce a novel cooperation scheme that facilitates cooperation and complements of branches to promote the overall learning ability. The cooperation scheme is constructed upon two principles. These principles promote each branch's learning ability, while also enabling them to uncover distinct patterns even with the same inputs.
    \item Through extensive experiments on our large-scale datasets, we demonstrate the effectiveness of the proposed method. 
    MBCnet has been deployed in image2product retrieval at Taobao app, and achieved obvious improvements .
\end{itemize}

\section{Related Work}
\textbf{Click-Through Rate (CTR) Prediction}: 
Capturing the complex feature interactions is a key to a successful CTR model.
Early works combine Logistic Regression (LR)~\cite{7154880} and Factorization Machine (FM)~\cite{5694074} to capture feature interactions. While these methods either rely much on handcrafted features or focus on lower-order feature interactions, showing limitations in learning complex feature patterns~\cite{YANG2022102853}. With the superior capacity of deep neural networks (DNN) for feature extraction, deep learning techniques have been widely examined in the context of CTR prediction~\cite{cheng2016wide,10.5555/3172077.3172127,wang2017deep}. 
For example,~\citet{cheng2016wide} combined shallow linear models and deep non-linear networks to promote the model's memorization and generalization ability for recommendation systems. \citet{10.5555/3172077.3172127} and~\citet{10.1145/3219819.3220023} combines the power of factorization machines and deep learning for feature learning. 
DeepIM~\cite{10.1145/3340531.3412077} effectively captures high-order interactions via a deep interaction machine module, serving as an efficient, exact high-order FM implementation.
\citet{wang2021dcn} designed Deep and Cross Network (DCN) which can more efficiently and explicitly learn certain bounded-degree feature interactions. They later introduced a mixture of low-rank expert network into DCN to make the model more practical in large-scale industrial settings~\cite{10.1145/3124749.3124754}. In FinalMLP~\cite{Mao023}, researchers observed that even two parallel MLP networks can achieve satisfied performance. They further proposed feature gating and interaction aggregation layers that can be easily plugged to make an enhanced two-stream MLP model. Some other CTR works investigate sequential user behavior modeling~\cite{zhou2018deep,zhou2019deep,ouyang2019deep}, cross-domain knowledge transfer~\cite{ouyang2020minet,li2019ddtcdr,9229522,chen2020towards,xu_CDAnet2024,9194523}, multi-task learning~\cite{ma2018modeling,tang2020progressive} and current popular large language models (LLMs) augmented CTR methods~\cite{geng2024breaking,lin2024clickprompt,muhamed2021ctr,li2023ctrl,huan2024exploring} for improved performance. These topics are beyond the scope of this paper, so we do not elaborate them here.

Existing works have studied different ways to model feature interactions and they usually contain two parallel branches for enhanced CTR performance. Our MBCnet also works in a multi-branch style, but it additionally contains a novel cooperation scheme that enables different branches to learn from each other and uncover various feature interaction patterns. Moreover, we design the EFGC branch, which can extensibly 
better incorporate various domain-driven knowledge and promote the model's memorization ability.

\textbf{Ensemble Learning}: Ensemble learning is a machine learning technique that aggregates two or more learners (\textit{e.g.}, regression models, neural networks) to produce better prediction performance~\cite{Yang_2022,9893798}. 
The fundamental idea is that single machine learning model suffers from various limitations (\textit{e.g.}, high variance, high bias, low accuracy). However, combining different models can address these limitations and achieve higher accuracy~\cite{Brown2010}.
For instance, bagging (\textit{e.g.}, random forest) reduces variance without increasing the bias, while boosting (\textit{e.g.}, XGBoost) reduces bias~\cite{10.1177/2332858420977208}. Generally speaking, an ensemble learner is more robust and able to perform better than the individual learners. In CTR prediction, different feature interaction techniques exhibit different modeling advantages. It is necessary to combine them together to better capture the complex patterns. Researchers have investigated the ensemble idea in CTR prediction from several aspects~\cite{10.1145/3041021.3054192}. \citet{10.1145/2648584.2648589} utilized GBDT for non-linear feature transformation and feed them to the LR for final prediction. While~\citet{10.1007/978-3-319-30671-1_4} used the FM embedding learned from sparse features to DNN. Researchers have found that simply replicating them cannot not yield satisfied performance. \citet{10.1145/3041021.3054192} investigated eight ensemble variants of GBDT, LR and DNN. They found that initializing the GBDT sample target with DNN's prediction score yielded the best performance in their business context. These methods do not work in an end-to-end style and require hard efforts on hand-craft tuning. Later, deep learning based ensemble CTR works~\cite{cheng2016wide,10.5555/3172077.3172127,wang2017deep,zhu_ensemble_ctr} are becoming more popular. For instance, 
CETnet~\cite{liu2024collaborative} focuses on the feature fusion of multiple models by a confidence based approach.
EnKD~\cite{zhu_ensemble_ctr} specifically studies knowledge distillation technique for model ensemble in order to learn light and accurate student prediction model. 
In addition, some multi-expert models~\cite{ma2018modeling,tang2020progressive} can also be regarded as using multi-branches for CTR modeling. They typically use several parallel MLP networks for generalized feature learning.

Compared to recent ensemble works~\cite{cheng2016wide,zhu_ensemble_ctr,Mao023,wang2017deep} and multi-expert models~\cite{ma2018modeling,tang2020progressive}, our MBCnet has one novel EFGC branch for extensible domain-driven feature interaction modeling and a completely different ensemble scheme. It combines multiple branches whose working scheme and architectures are different, while multi-expert models usually have the same expert architectures and rely on parameter initialization for learning different pattern. Furthermore, the cooperation scheme explicitly advocates the branches to learn from each other and uncover different feature patterns even under the same supervision.

\begin{figure*}[t]
\centering
\includegraphics[width=14.5cm]{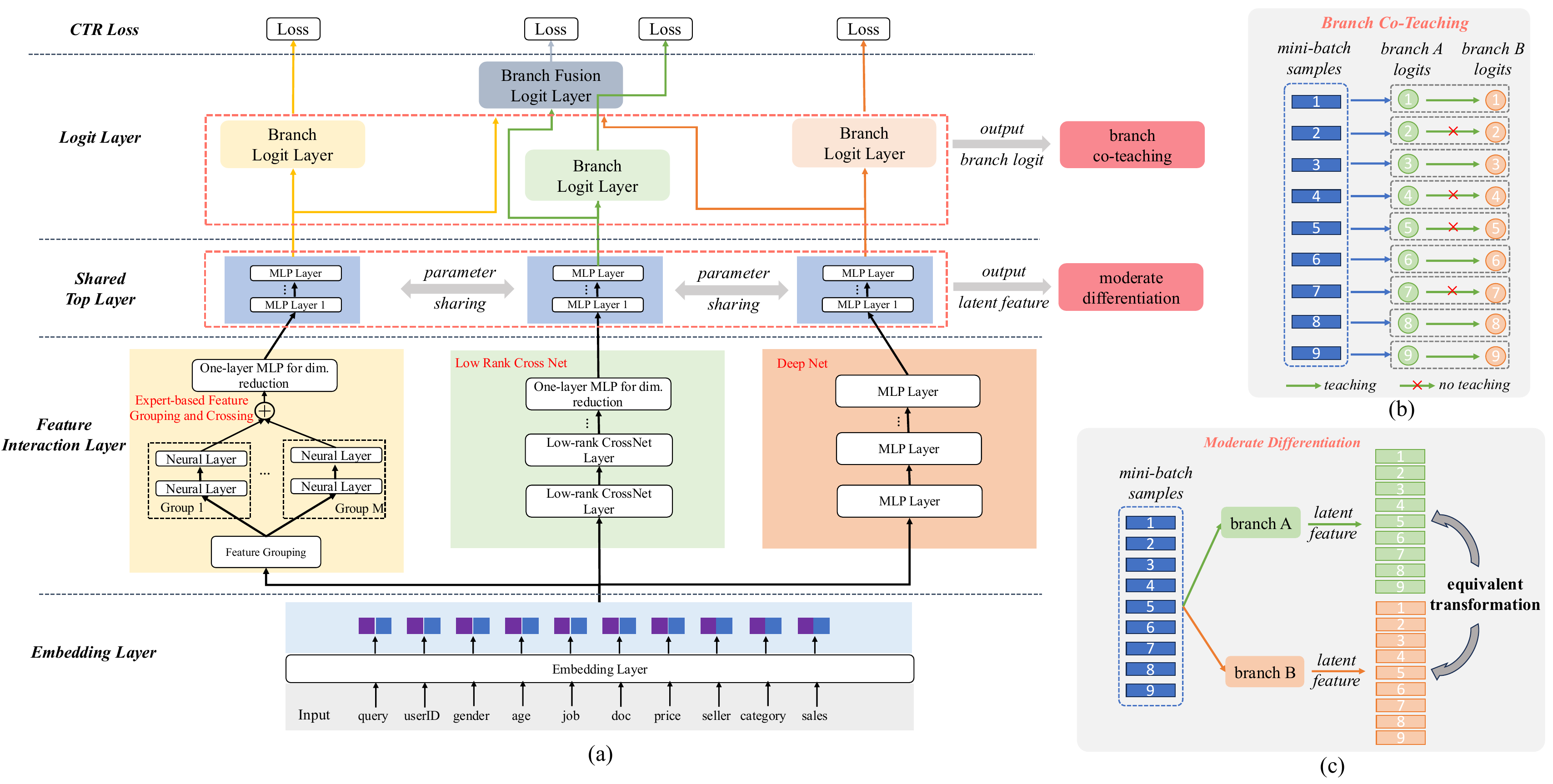}
\vspace{-14pt}
\caption{(a) shows the architecture of our multi-branch cooperation networks (MBCnet). It consists of three different branches: the EFGC branch, the low-rank Cross Net branch and the Deep net branch. To enlarge the model's capability, there is a novel multi-branch cooperation scheme based on two principles: \textbf{\textit{branch co-teaching}} and \textbf{\textit{moderate differentiation}}. (b) illustrates the branch co-teaching approach between any two of the three branches. (c) demonstrates that latent features from any two branches undergo equivalent transformation.
} 
\label{figure:model_architecture}
\vspace{-10pt}
\end{figure*}
\section{Method}
\subsection{Overview}
In CTR prediction, given the raw input feature $\boldsymbol{x}=\{\boldsymbol{x_{1}}, \boldsymbol{x_{2}},..., \boldsymbol{x_{M}}\}$ with $M$ feature fields and user feedback label $\boldsymbol{y}\in \{0,1\}$, we aim to learn a mapping function $\mathcal{F}: \boldsymbol{x}\rightarrow \boldsymbol{y}$ to predict user behaviors for online services. Usually, $\mathcal{F}$ denotes different deep neural networks (\textit{e.g.}, DCNv2~\cite{wang2021dcn}) and $\boldsymbol{x_i}$ denotes the $i$-th feature field that can be \textit{categorical}, \textit{multi-valued}, or \textit{numerical} feature of users or items. 
Especially in industrial scenarios, the number of feature fields $M$ can be in hundreds or even thousands, bringing great challenges of modeling feature interactions.

MBCnet comprises three feature interaction branches: the proposed EFGC branch, the low-rank Cross Net branch, and the Deep branch. It also includes an innovative multi-branch cooperation scheme. The general architecture is given in Figure~\ref{figure:model_architecture}. Details are provided in subsequent sections.

\subsection{Extensible Feature Grouping and Crossing}
The EFGC branch is designed to improve the model's memorization capability for intended interactions of specific feature fields, guided by domain-driven knowledge. It consists of two main components: the embedding layer, and the feature grouping and crossing module.   

\textbf{Embedding Layer}: The embedding layer aims to encode raw input feature $\boldsymbol{x}=\{\boldsymbol{x_1}, \boldsymbol{x_2},..., \boldsymbol{x_{M}}\}$ as embedding vectors. 
Each field has specific meaning such as \textit{userID}, \textit{age} and \textit{gender}. Let $E_{i}$ be the $i$-th embedding function of $\boldsymbol{x_{i}}$, the embedding layer is written as:
\begin{align}
\label{eq:embedding_layer}
    \{\boldsymbol{e_{1}}, \boldsymbol{e_{2}}, ..., \boldsymbol{e_{M}}\} = \{E_{1}(\boldsymbol{x_{1}}), E_{2}(\boldsymbol{x_{2}}), ..., E_{M}(\boldsymbol{x_{M}})\}
\end{align}
where $\boldsymbol{e}_{i}$ is the embedding vector of $\boldsymbol{x_{i}}$.

\textbf{Feature Grouping}: In CTR prediction, different feature fields have different physical meanings, and their interactions are crucial for personalized ranking. For instance, in e-commence search, combining user profile feature fields (\textit{e.g., userID, age}) and item profile feature fields (\textit{e.g., itemID, price, sales volume}) can reveal a user's general preferences regarding item price and quality. Conversely, interactions like \textit{item title} and \textit{item category} could show less contributions to personalized ranking. \textit{In essence, different feature field interactions have varying degrees of influence on CTR prediction.} The common practice of concatenating $\boldsymbol{e}_{i}$ in Eq.~\ref{eq:embedding_layer} may fail to highlight certain specific interactions and introduce irrelevant feature patterns into subsequent deep modules.

\begin{table}[]
\centering
\caption{The designed feature groups of EFGC branch in our image2product search.}
\vspace{-8pt}
\label{table:EFGC_groups}
\renewcommand{\arraystretch}{1.5}
 \setlength{\tabcolsep}{0.5mm}{ 
  \scalebox{0.6}{
\begin{tabular}{ccc}
\hline
group & feature fields                       & intention of feature interactions                                          \\ \hline
1     & \textit{query image, item image}     & visual relevance between query and item            \\ \hline
2     & \textit{query image, item attributes}   & relevance between visual query and item attributes \\ \hline
3     & \textit{user profile, item image}    & user preference on visual patterns of images        \\ \hline
4     & \textit{user profile, item attributes}  & generalized user preference on item attributes                 \\ \hline
5     & \textit{userID, itemID}  & personalized user preference on items                 \\ \hline
6     & \textit{queryClusterID, docClusterID}  & matching degree between clustered query and item images                 \\ \hline
7     & \textit{queryClusterID attributes, docClusterID attributes} & non-visual relevance between query and item        \\ \hline
\end{tabular}
}}
\vspace{-10pt}
\end{table}
It is crucial to organize feature fields into distinct groups and perform feature crossing within these specific groups, driven by domain knowledge. To achieve this, we can follow three main steps: 1) \textit{Identify Desired Matching Patterns}: Determine the types of matching patterns that need to be emphasized based on the problem domain.
2) \textit{Group Feature Fields}: Select and group the feature fields in a way that aligns with the identified matching patterns.
3) \textit{Integrate Group Embeddings into Subsequent Layers}: Input the embeddings from each group into the following layers to facilitate the learning of feature interactions.
For instance, in step 1 of our image2product search, we aim to emphasize the visual relevance between query image and item image, while personalizing the results based on each user's preferences on item attributes. Step 2 is formulated as:
\begin{align}
    \boldsymbol{e^{\text{Group}}_{i}} = \boldsymbol{e_{x}}\oplus,...,\oplus \boldsymbol{e_{y}}
\end{align}
where $\boldsymbol{e^{\text{Group}}_{i}}$ denotes $i$-th group that combines embeddings of specific feature fields. $\oplus$ is the concatenation operation. Detailed designed groups in our image2product search are listed in Table~\ref{table:EFGC_groups}. In this table, query and item image are encoded as vectors by deep networks. The \textit{user profile} includes fields such as \textit{userID, age, job, gender, consumption level} and other personalized profiles. The \textit{item attributes} encompasses fields like \textit{itemID, price, sales volume, delivery location, brand} and other relevant information. The \textit{queryClusterID} and \textit{docClusterID} mean the clustering ID of query images and item images.
The \textit{queryClusterID attributes} includes features such as \textit{query image category}, \textit{average price of query related items}.

\textbf{Feature Crossing}: After feature grouping, we can perform feature crossing with subsequent MLP layers as follows:
\begin{align}
    \boldsymbol{h_{i}^{\text{EFGC}}} = f_{i}^{\text{EFGC}}(\boldsymbol{e^{\text{Group}}_{i}})
\end{align}
where $\boldsymbol{h_{i}^{\text{EFGC}}}$ denotes the output of $i$-th group feature crossing. $f_{i}^{\text{EFGC}}$ is a non-linear MLP network. Finally, as shown in Figure~\ref{figure:model_architecture}, the output of our EFGC branch, denoted as $\boldsymbol{h^{\text{EFGC}}}$, is formulated as:
\begin{align}
    \boldsymbol{h^{\text{EFGC}}} = \mathbb{FC}(\boldsymbol{h_{i}^{\text{EFGC}}}\oplus,...,\oplus \boldsymbol{h_{N_{g}}^{\text{EFGC}}})
\end{align}
where $N_{g}$ is the number of feature groups and $\mathbb{FC}$ means one fully-connected layer to reduce the output dimension. 

\underline{\textbf{\emph{Remark}}}: Through splitting feature fields into groups and performing feature crossing within these groups, \textit{EFGC branch enhances the model's ability to memorize specific patterns and avoid undesired feature crossing.} \textit{Additionally, it is extensible to include more types of feature groups as long as desired matching patterns are given.} The proposed EFGC is extensible and easily applied to other fields by using the designed feature groups in Table~\ref{table:EFGC_groups}.

We note feature engineering techniques also use specific feature fields to create new features, such as the count of clicks between \textit{userID} and \textit{brand}. Both feature engineering and EFGC employ domain-driven knowledge. However, the difference is feature engineering creates raw input features, whereas EFGC learns feature crossing within designed groups, enabling the automatic discovery of patterns among feature fields within groups. They are not mutually exclusive and can be used together in practice.

\subsection{Deep Net and Low Rank Cross Net}
\textbf{Deep Net}: Deep neural networks, composed of stacked non-linear MLP layers, have the capability to implicitly learn feature interactions~\cite{cheng2016wide,10.1145/3219819.3220023,10.5555/3172077.3172127,wang2021dcn,10.1145/3124749.3124754,Mao023}. The input of Deep Net is typically formed by concatenating all embedding fields as:
\begin{align}
    \boldsymbol{e^{\text{Deep}}} = \boldsymbol{e_{1}}\oplus \boldsymbol{e_{2}}\oplus,...,\oplus \boldsymbol{e_{M}}
\end{align}
Let $f^{\text{Deep}}$ be the stacked non-linear MLP layers of deep Net branch. The output of Deep Net branch, denoted as $\boldsymbol{h^{\text{Deep}}}$, is written as:
\begin{align}
    \boldsymbol{h^{\text{Deep}}} = f^{\text{Deep}}(\boldsymbol{e^{\text{Deep}}})
\end{align}

\textbf{Low Rank Cross Net}: Despite the effectiveness of Deep Net in modern applications, it often struggles with capturing feature interactions, especially higher-order ones~\cite{wang2021dcn}. 
The popular low rank cross net(\textit{i.e.}, CrossNetV2) maps feature interactions in low-rank space and employs a mixture of experts architecture to improve its expressiveness. 
It is cost-efficient and expressive for feature interaction learning. The input of low-rank cross net branch is also the concatenation of all field embeddings, which means $\boldsymbol{e^{\text{Cross}}}=\boldsymbol{e^{\text{Deep}}}$.Then the resulting low-rank cross net layer is formulated as:
\begin{align}
    \boldsymbol{h^{\text{Cross}}} = \mathbb{FC}(f^{\text{Cross}}(\boldsymbol{e^{\text{Cross}}}))
\end{align}
where $\mathbb{FC}$ denotes one MLP layer for dimension reduction. $f^{\text{Cross}}$ denotes a mixture of low-rank cross network blocks.
Due to the page limit, we do not elaborate it in detail here. 
For a detailed formulation, please refer to Appendix~\ref{appendix:dcnv2} or~\cite{wang2021dcn}. After the feature interaction learning of each branch, we further include a shared top layer $f_{\theta}(\cdot)$ to reduce parameters and regularize training. The output latent features of different branches are formulated as:
\begin{align}
    \boldsymbol{z^{\text{EFGC}}}=f_{\theta}(\boldsymbol{h^{\text{EFGC}}}),~\boldsymbol{z^{\text{Deep}}}=f_{\theta}(\boldsymbol{h^{\text{Deep}}}),~\boldsymbol{z^{\text{Cross}}}=f_{\theta}(\boldsymbol{h^{\text{Cross}}})
\end{align}
where $\boldsymbol{z^{\text{EFGC}}}$, $\boldsymbol{z^{\text{Deep}}}$ and $\boldsymbol{z^{\text{Cross}}}$ are also the input of logit layer.


\subsection{Multi-branch Cooperation}
\subsubsection{\textbf{Branch Co-teaching}}
Different branches of MBCnet possess distinct advantages and inherently focus on modeling different patterns. These patterns, in turn, are represented by the training samples. As a result, the learning capabilities of different branches can vary significantly across different samples. On particular samples, if a robust branch has been effectively trained, it can assist a comparatively weak branch in enhancing its performance. We formulate this collaboration idea as "\textit{\textbf{branch co-teaching}}" objective function, whereby given specific samples, the strong branch provides guidance to the weak branch in training. 

\textbf{Sample Selection by Disagreement}: To perform the above co-teaching, we first need to identify on which samples that one branch exhibits strong prediction performance while the other one shows weak performance. In other words, two branches have disagreement predictions on these samples.
In CTR prediction, it is naturally to use binary cross entropy (BCE) loss to measure whether a model has accurate prediction on one sample. 
To be specific, we denote the predicted click probability of one branch as:
\begin{align}
    \boldsymbol{p^{i}}=\sigma(f_{\phi}^{i}(\boldsymbol{z^{i}}))
\end{align}
where $i\in \{\text{EFGC, Deep, Cross}\}$, $\sigma(\cdot)$ denotes the sigmoid function. $f_{\phi}^{i}(\cdot)$ and $\boldsymbol{z^{i}}$ denote the logit layer and latent feature of branch $i$, respectively.
Therefore on one particular sample, we can identify strong and weak branches as:
\begin{align}
\label{eq:p_j_strong}
    (\boldsymbol{p^j} \text{ is strong},\boldsymbol{p^i} \text{ is weak})~\text{\textit{if}}~
    \begin{cases}
        &BCE(\boldsymbol{p^{j},\boldsymbol{y}})<-\log(0.5) \\ &BCE(\boldsymbol{p^{i},\boldsymbol{y}})>-\log(0.5)
    \end{cases}
\end{align}
\begin{align}
\label{eq:p_i_strong}
    (\boldsymbol{p^i} \text{ is strong},\boldsymbol{p^j} \text{ is weak})~\text{\textit{if}}~
    \begin{cases}
        &BCE(\boldsymbol{p^{i},\boldsymbol{y}})<-\log(0.5) \\ &BCE(\boldsymbol{p^{j},\boldsymbol{y}})>-\log(0.5)
    \end{cases}
\end{align}
where $BCE$ means binary cross entropy loss. Since CTR prediction is usually regarded as a binary classification task, we employ $-\log(0.5)$ as a threshold to assess the learning progress of each branch\footnote{The prediction probabilities may be affected by the data distribution, we experimented with various thresholds and found $-\log(0.5)$ generally did the best.}. When the loss value is less than $-\log(0.5)$, we consider the branch on this samples is learned effectively. Otherwise, we consider the branch is not learned well.

\textbf{Branch Co-teaching Objective Formulation}: Given the above defined strong and weak branches, we take the prediction of strong branches as soft label to guide the learning of weak branches on selected samples. The co-teaching loss is written as follows:
\begin{align}
\label{eq:co_teaching}
\small
\mathcal{L}_{BCT}= -\frac{1}{C}\sum_{i}\sum_{j\neq i} [I^{ij}\text{SG}(\boldsymbol{p^i})\log(\boldsymbol{p^{j}})+I^{ji}\text{SG}(\boldsymbol{p^j})\log(\boldsymbol{p^{i}})]
\end{align}
where $i,j\in \{\text{EFGC}, \text{Deep}, \text{Cross}\}$. SG means stop gradient. $I^{ij}\in \{0,1\}$ is the indicator that only equals 1 when $\boldsymbol{p^i}$ is strong and $\boldsymbol{p^j}$ is weak shown in Eq.~\ref{eq:p_i_strong}. Similarly, $I^{ji}\in\{0,1\}$ is the indicator that equals 1 only when $\boldsymbol{p^j}$ is strong and $\boldsymbol{p^i}$ is weak shown in Eq.~\ref{eq:p_j_strong}. 
In other cases, both $I^{ij}$ and $I^{ji}$ equal 0, which means we will not perform co-teaching optimization on those samples.
\textit{This suggests that when two learners do not produce significantly different predictions, transferring teaching guidance may be unnecessary. It allows both learners the flexibility to discover and capture their own patterns.} We conducted an experiment to demonstrate this idea in Section~\ref{sec:how_multi_branch}.
$C$ is the number of non-zero values calculated from $I^{ij}$ and $I^{ji}$, indicating how many disagreement pairs are used in $\mathcal{L}_{BCT}$.

\textit{\textbf{\underline{Remark}}}: The main idea of this bidirectional branch co-teaching objective is that the teaching guidance is only conducted on selected samples where two branches show disagreed predictions. It is one important difference between this approach and the knowledge distillation in~\cite{zhu_ensemble_ctr} which performs full sample supervision between teachers and students. Given specific samples, learning abilities of different branches vary and are gradually improved, which suggests that the branches cannot offer appropriate supervision for all samples, especially in the early training stage. Employing supervision on all samples could easily introduce noisy supervision, and this is why EnKD~\cite{zhu_ensemble_ctr} uses pre-trained teachers and temperature scaling tricks to mitigate the effects of such noise. In contrast, our branch co-teaching operates in an end-to-end manner, focusing solely on samples with disagreeing predictions to minimize supervision noise.

\subsubsection{\textit{\textbf{Moderate Differentiation}}}
In MBCnet, multiple feature interaction branches make predictions for the same input sample. When these branches are supervised on the same sample, they have risk to learn identical feature interaction patterns. This redundancy can significantly hinder the model's capacity to explore a variety of feature interactions. To enhance the model's robustness, it is better to enable distinct branches to learn differentiated latent features, encouraging them to capture diverse interaction patterns even on the same sample. \textit{However, excessive differentiation between these latent features may disrupt the consistency of shared patterns across branches, potentially harming model performance}. In this context, we propose "\textit{\textbf{\textit{moderate differentiation}}}" for the latent features $\boldsymbol{z^i}$ across branches. This approach promotes the exploration of diverse interaction patterns while ensuring a balance between differentiation and consistency.

\textbf{Equivalent Transformation of Latent Features}:
To achieve moderate differentiation between any two branches, we assume that their latent features satisfy equivalent transformation~\cite{hefferon2018linear}. This indicates these two latent features have relations but also learning flexibility to ensure differentiation~\cite{chen2020towards}. 
Specifically, if $\boldsymbol{z^i}\in \mathbb{R}^{1\times d}$ and $\boldsymbol{z^j}\in \mathbb{R}^{1\times d}$ are the latent features from two distinct branches on the same sample, they satisfy the equivalent transformation as\footnote{Note that we use an simplified version of equivalent transformation as in~\cite{chen2020towards}}:
\begin{align}
\boldsymbol{z^i}=\boldsymbol{z^j} W^{ji},~\boldsymbol{z^j}=\boldsymbol{z^i} W^{ij},~s.t.~W^{ji} W^{ij}=I_{d}
\end{align}
where $W^{ji}, W^{ij}\in \mathbb{R}^{d\times d}$ are two learnable matrices and they are inverses of each other. To avoid feature collapse where multiple $\boldsymbol{z^i}$ collapse to one point after transformation, we employ orthogonal transformation, a specification of equivalent transformation. This means $W^{ji}$ and $W^{ij}$ are the transpose to each other, \textit{i.e.}, $W^{ij} = (W^{ji})^{T}$. It is important to point out that the above constraint is not equivalent to $\boldsymbol{z^i}$ and $\boldsymbol{z^j}$ are orthogonal. The orthogonal transformation is a special case of equivalent transformation, and enables $\boldsymbol{z^i}$ and $\boldsymbol{z^j}$ to maintain a moderate-level differentiation. It has an advantage of preserving inner product of vectors, namely it can keep similarities of vectors after transformation.

\textbf{Moderate Differentiation Regularization}: Taking the above analysis into consideration, we can write the regularization loss for moderate differentiation as:
\begin{align}
\label{eq:moderate_diff}
    \mathcal{L}_{MDR} = \frac{1}{K(K-1)}\sum_{i}\sum_{j\neq i} ||\boldsymbol{z^i} W^{ij} - \boldsymbol{z^j}||_{F}^2 + ||\boldsymbol{z^i} W^{ij} (W^{ij})^{T} - \boldsymbol{z^i}||_{F}^2
\end{align}
where the first term indicates the transformation relationship between $\boldsymbol{z^i}$ and $\boldsymbol{z^j}$, while the second term shows the constraint of orthogonal transformation on $W^{ij}$.
\textit{This loss not only enhances the model's capability but also facilitates the learning process by maintaining relevant feature representations across branches.}

\subsubsection{\textbf{\textit{Branch Fusion and Training Objective Function}}}
We perform branch fusion to incorporate their knowledge and make better behavior predictions. Specifically, given latent feature of three branches, we have:
\begin{align}
\boldsymbol{z^{\text{fusion}}}=Avg\_Pool(\boldsymbol{z^\text{EFGC}}, \boldsymbol{z^\text{Deep}}, \boldsymbol{z^\text{Cross}})
\end{align}
where $\boldsymbol{z^{\text{fusion}}}$ is the fused latent feature of different branches. $Avg\_Pool$ means average pooling of latent features in feature dimension.
Then, the training objective for CTR prediction is formulated as:
\begin{align}
    \mathcal{L}_{CTR} = \sum_{i} BCE(\sigma(f_{\phi}^{i}(\boldsymbol{z^i})), \boldsymbol{y})
\end{align}
where $i\in \{\text{fusion},\text{EFGC},\text{Deep},\text{Cross} \}$ and $BCE$ is the binary cross entropy loss. $f_{\phi}^i(\cdot)$ indicates the logit layer. $\sigma(\cdot)$ denotes the sigmoid function. It is also worthwhile to mention that BCE loss is one choice for CTR prediction, it can be replaced with other CTR loss functions. To sum up, the overall objective function of MBCnet is defined as:
\begin{align}
    \mathcal{L} = \mathcal{L}_{CTR}+\alpha*\mathcal{L}_{BCT}+\beta*\mathcal{L}_{MDR}
\end{align}
where $\alpha$ and $\beta$ are two hyper-parameters to weight the importance of loss terms. 
Notice that we use an arbitrary sample to succinctly demonstrate above objective function with simple symbols. When using the objective function in practice, we compute the expectation of loss values of mini-batch samples during training.

\textit{\textbf{\underline{Discussion}}}: Popular CTR works~\cite{cheng2016wide,wang2021dcn,10.5555/3172077.3172127,Mao023} investigate various feature interaction techniques for improved performance. 
However, each technique offers its own advantages, and relying on just one may not adequately capture the patterns in complex datasets, especially in industrial scenarios with enormous users and items.
Meanwhile, these methods usually simply combine them with prediction voting or latent feature concatenation. By contrast, the proposed MBCnet advocates different branches to \textbf{explicitly learn from each other} and cooperate together to form a more powerful model, which is a new idea in CTR ranking. Compared to recent research~\cite{zhu_ensemble_ctr,liu2024collaborative} that also leverages supervision from other models, MBCnet implements a distinct and more advanced learning framework.
The comparison results presented in the following sections will highlight this difference.

\textbf{We have also made some theoretical understanding of our cooperation scheme.} In ensemble learning~\cite{zhou2025ensemble} and co-training~\cite{10.5555/3327757.3327944}, usually two or more learners with different parameter initialization are used to keep learner divergence. This divergence helps the whole model to filter different types of errors and improve the model generalization ability. However, these learners would gradually converge to be close to each other as the training epochs increase because of rough learner alignment, and lose learner divergence for generalization. \textbf{The cooperation scheme in MBCnet provides an explicit and effective way to keep the divergence in training.} First, the branch co-teaching transfers supervision knowledge from strong branches to weak branches on disagreement samples. In other words, within the whole training epochs, two branches can learn from each other and always select disagreement data for training, so the divergences of branches will be always maintained. This can also be supported by the visualization in Figure~\ref{figure:why_mbc} (d). Second, the moderate differentiation provides a good way to ensure learners to have a moderate-level learning divergence, and avoids both too large or too small divergence for this CTR learning.

\section{Experiment and Analysis}

\subsection{Experiment Setup}
\subsubsection{Datasets}
Since there are no public datasets with hundreds of rich feature fields, we conduct experiments on two industrial datasets: Pailitao-12month and Pailitao-24month. Both are collected from our image2product search, also known as ``Pailitao'' at Taobao. Pailitao-12month and Pailitao-24month contain user search behaviors ranging from 2023-06-01 to 2024-05-31 and 2022-06-01 to 2024-05-31, respectively. For both datasets, there are various feature fields, including query image feature, user profiles, user statistical features, item image feature, item attributes, item statistical features and some other designed clustering and crossing features. The dataset statistics are given in Table~\ref{table:dataset}, which show good resource to verify our method for the industrial track submission. \textbf{All experimental models utilize the same input features to ensure a fair comparison of their modeling differences.}

\begin{table}[]
\centering
\caption{The statistics of used industrial datasets.}
\vspace{-8pt}
\label{table:dataset}
\renewcommand{\arraystretch}{0.9}
 \setlength{\tabcolsep}{1.0mm}{ 
  \scalebox{0.9}{
\begin{tabular}{ccc}
\hline
Dataset            & Pailitao-12month & Pailitao-24month \\ \hline
\#user            & 480,028,940      & 528,665,908      \\
\#item            & 974,333,635      & 1,355,865,204    \\
\#feature\_fields & 192              & 192              \\
\#feature\_dim  & 2780             & 2780             \\
\#samples         & 203.99 billion   & 395.99 billion   \\
density            & 4.36e-5\%        & 5.52e-4\%        \\ \hline
\end{tabular}
}}
\vspace{-10pt}
\end{table}
\subsubsection{Baselines}
We leverage different strong algorithms for comparison including \textbf{single branch methods} and \textbf{branch ensemble} methods. The \textbf{single branch methods} are as follows. 1) \textbf{DNN} is a deep multi-layer perception model, 2) \textbf{CrossNetV2}~\cite{wang2021dcn} is a low-rank version of cross net~\cite{10.1145/3124749.3124754} to explicitly learn high-order feature interactions.  
The \textbf{branch ensemble methods} are as follows. 
3) \textbf{Wide\&Deep}~\cite{cheng2016wide} combines a wide set of cross-product feature
transformations and deep neural networks.
4) \textbf{DCNv2}~\cite{wang2021dcn} contains mixture-of-expert based low-rank feature crossing and DNN to learn bounded-degree feature interactions. 
5) \textbf{MMOE+}~\cite{ma2018modeling} originally is a multi-task learning model. We employ its mixture-of-expert encoder to work as a baseline with multiple-expert ensemble. 
6) \textbf{EnKD}~\cite{zhu_ensemble_ctr} studies knowledge distillation technique for model ensemble. 
7) \textbf{FinalMLP}~\cite{Mao023} designs feature gating and interaction aggregation layers for an enhanced CTR model.

\begin{table}[]
\centering
\caption{The AUC and LogLoss comparison results of different models on our two datasets. Best results are in bold and the most competitive public baseline results are underlined.} 
\vspace{-8pt}
\label{table:offline_overall_comparison}
\renewcommand{\arraystretch}{0.9}
 \setlength{\tabcolsep}{1.0mm}{ 
  \scalebox{1.0}{
\begin{tabular}{cc|cc|cc}
\hline
\multicolumn{2}{c|}{Dataset}                                & \multicolumn{2}{c|}{Pailitao-12month}                                                                                                     & \multicolumn{2}{c}{Pailitao-24month}                                                                                                      \\ \hline
\multicolumn{2}{c|}{Model}                                  & AUC$\uparrow$                                                       & LogLoss$\downarrow$                                                 & AUC$\uparrow$                                                       & LogLoss$\downarrow$                                                 \\ \hline
\multicolumn{1}{c|}{\multirow{2}{*}{Single}}   & DNN        & 0.7423                                                              & 0.2894                                                              & \underline{0.7569}                                                        & \underline{0.2829}                                                        \\
\multicolumn{1}{c|}{}                          & CrossNetV2 & 0.7437                                                              & 0.2917                                                              & 0.7555                                                              & 0.2835                                                              \\ \hline
\multicolumn{1}{c|}{\multirow{5}{*}{Ensemble}} & Wide\&Deep & 0.7392                                                              & 0.2911                                                              & 0.7513                                                              & 0.2863                                                              \\
\multicolumn{1}{c|}{}                          & DCNv2      & \underline{0.7461}                                                        & 0.2872                                                              & 0.7559                                                              & 0.2837                                                              \\
\multicolumn{1}{c|}{}                          & MMOE+      & 0.7400                                                              & 0.2906                                                              & 0.7544                                                              & 0.2837                                                              \\
\multicolumn{1}{c|}{}                          & EnKD       & 0.7450                                                              & \underline{0.2871}                                                        & 0.7541                                                              & 0.2834                                                              \\
\multicolumn{1}{c|}{}                          & FinalMLP   & 0.7427                                                              & 0.2891                                                              & 0.7556                                                              & 0.2838                                                              \\ \hline
\multicolumn{1}{c|}{\multirow{2}{*}{Ours}}     & EFGC       & 0.7490                                                              & 0.2880                                                              & 0.7600                                                              & 0.2825                                                              \\ \cline{2-6} 
\multicolumn{1}{c|}{}                          & MBCnet     & \textbf{\begin{tabular}[c]{@{}c@{}}0.7522\\ (+0.61\%)\end{tabular}} & \textbf{\begin{tabular}[c]{@{}c@{}}0.2866\\ (-0.05\%)\end{tabular}} & \textbf{\begin{tabular}[c]{@{}c@{}}0.7642\\ (+0.72\%)\end{tabular}} & \textbf{\begin{tabular}[c]{@{}c@{}}0.2800\\ (-0.29\%)\end{tabular}} \\ \hline
\end{tabular}
}}
\vspace{-12pt}
\end{table}
\subsubsection{Parameter Settings}
In our experiments, we split the datasets by chronological order, with the last seven-day data as validation and test set, and the rest is taken as the train set. For all methods, we set the latent dimension of last feature learning layer as 128. The validation performance is set as early-stop condition in training. In MBCnet, $\alpha$ and $\beta$ both equal 0.1 by hyper-parameter searching. In EFGC branch, hidden units of each group crossing are [1024,128]. In CrossNet branch, the number of experts is 2 and each expert has 2 layers with low rank dimension as 16. 
The hidden units of dimension reduction layer in EFGC and CrossNet are both 512. In Deep branch, the hidden units are [2048,1024,512,512,512]. Hidden units of the shared top layer are [512,256,128].
Hyper-parameters of baselines are searched to ensure each achieves its best performance. 
\textbf{Core codes are available in the Appendix.}

\subsection{Overall Comparison}
\subsubsection{\textbf{Offline Comparison}}
\label{sec:offline_comparison}
We make model comparisons with recent competitive baselines on our two large-scale industrial datasets. The results are given in Table~\ref{table:offline_overall_comparison}.
From this table, we can observe that: 1) our MBCnet achieves obvious improvement over recent competitive baselines. It has a 0.61\% and 0.72\% AUC increase over the most competitive model on two datasets, which are impressive improvements in large-scale industrial data scenarios. 
2) Even when solely using our EFGC model, it can achieve superior model performance over other baselines, demonstrating its effectiveness. EFGC is a simple but effective design, and can be regarded as a variant of DNN branch. It has advanced capabilities for reducing redundant feature interactions and effectively learning domain-driven knowledge.  
3) On Pailitao-24month dataset, single DNN outperforms other baselines. This trend is commonly observed in large-scale industrial settings, where the enormous data can overwhelm many modeling techniques. Even in this context, MBCnet still achieves superior improvements, demonstrating the effectiveness of our approach that leverages cooperative multi-branches to predict user behaviors.

\begin{table}[]
\centering
\caption{The comparison results when deploying MBCnet in our production environment. AUC is the offline evaluation metric. CTR, deal number and GMV are online A/B test metrics. Some values of the base model are denoted as $\star$ to obey the company's regulations.} 
\vspace{-8pt}
\label{table:product_results}
\renewcommand{\arraystretch}{1.0}
 \setlength{\tabcolsep}{0.8mm}{ 
  \scalebox{0.9}{
\begin{tabular}{cccccc}
\hline
Model  & AUC    & CTR                 & \begin{tabular}[c]{@{}c@{}}deal number \\ by query\end{tabular} & \begin{tabular}[c]{@{}c@{}}deal number \\ by user\end{tabular} & GMV    \\ \hline
Base   & 0.7559 & 10.37\%             & $\star$                                                               & $\star$                                                              & $\star$      \\ \hline
MBCnet & \textbf{0.7642} & \textbf{10.46\%(+0.09point)} & \textbf{+1.49\%}                                                         & \textbf{+1.18\%}                                                         & \textbf{+1.62\%} \\ \hline
\end{tabular}
}}
\vspace{-8pt}
\end{table}

\begin{figure}[t]
\centering
\begin{minipage}[t]{0.22\textwidth}
\centering
\includegraphics[width=\textwidth]{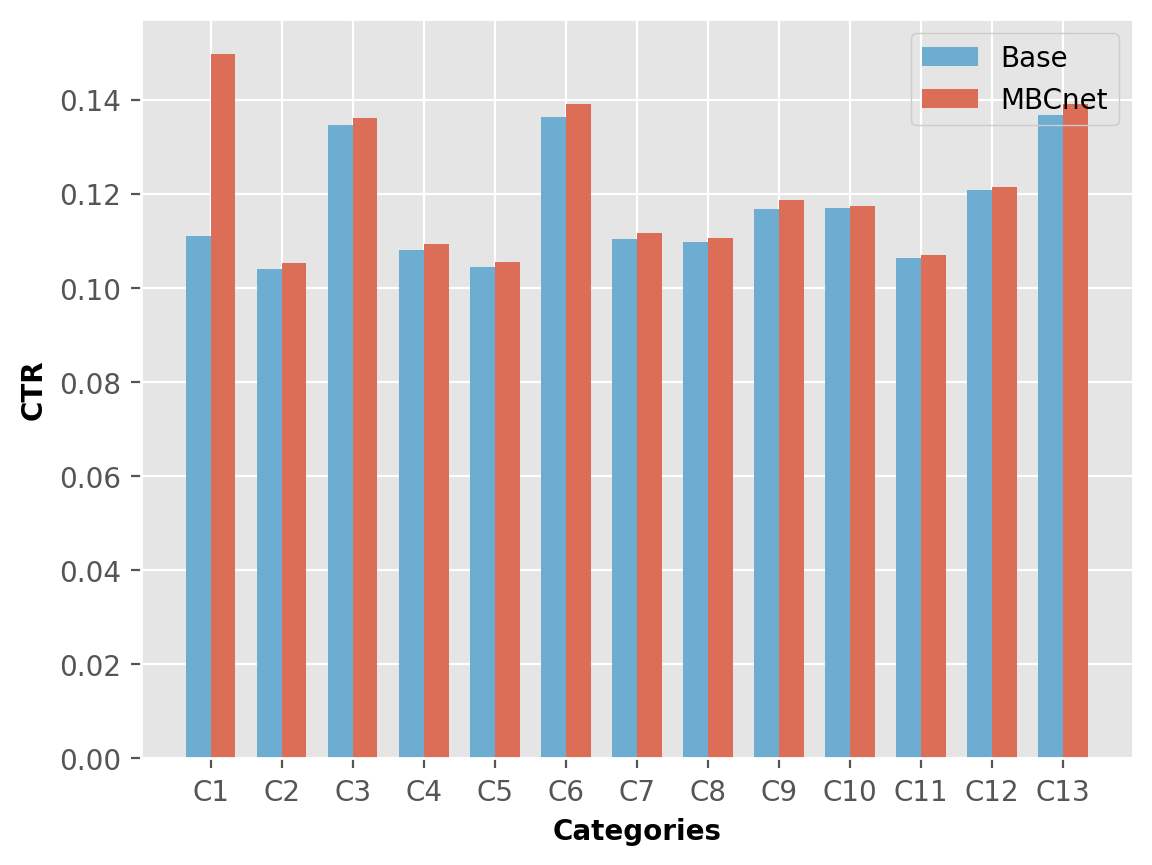}
\vspace{-20pt}
\caption*{{\small(a) CTR under Query Category}}
\end{minipage}
\begin{minipage}[t]{0.22\textwidth}
\centering
\includegraphics[width=\textwidth]{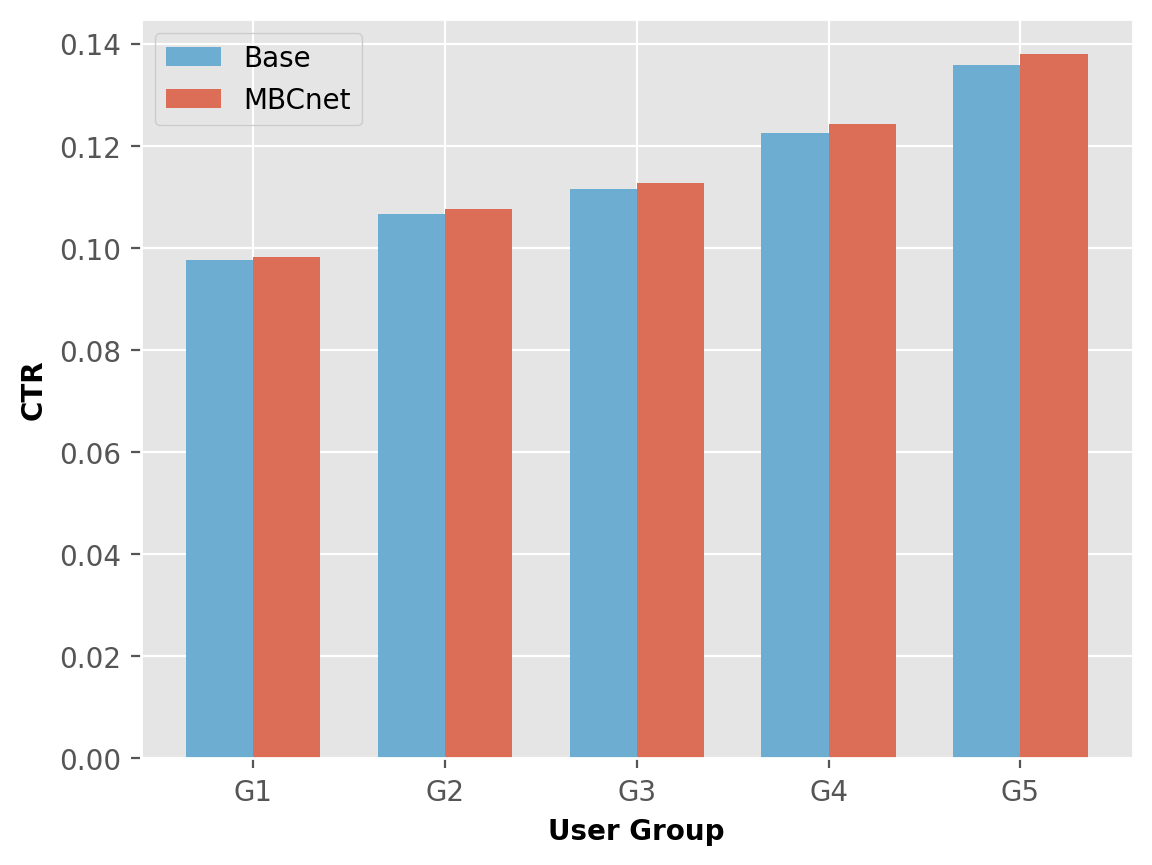}
\vspace{-20pt}
\caption*{\small{(b) CTR under User Group}}
\end{minipage}
\vspace{-8pt}
\caption{Online CTR comparison under different query categories and user groups.}
\label{figure:online_AB}
\vspace{-10pt}
\end{figure}

\begin{figure*}[t]
\centering
\begin{minipage}[t]{0.22\textwidth}
\centering
\includegraphics[width=\textwidth]{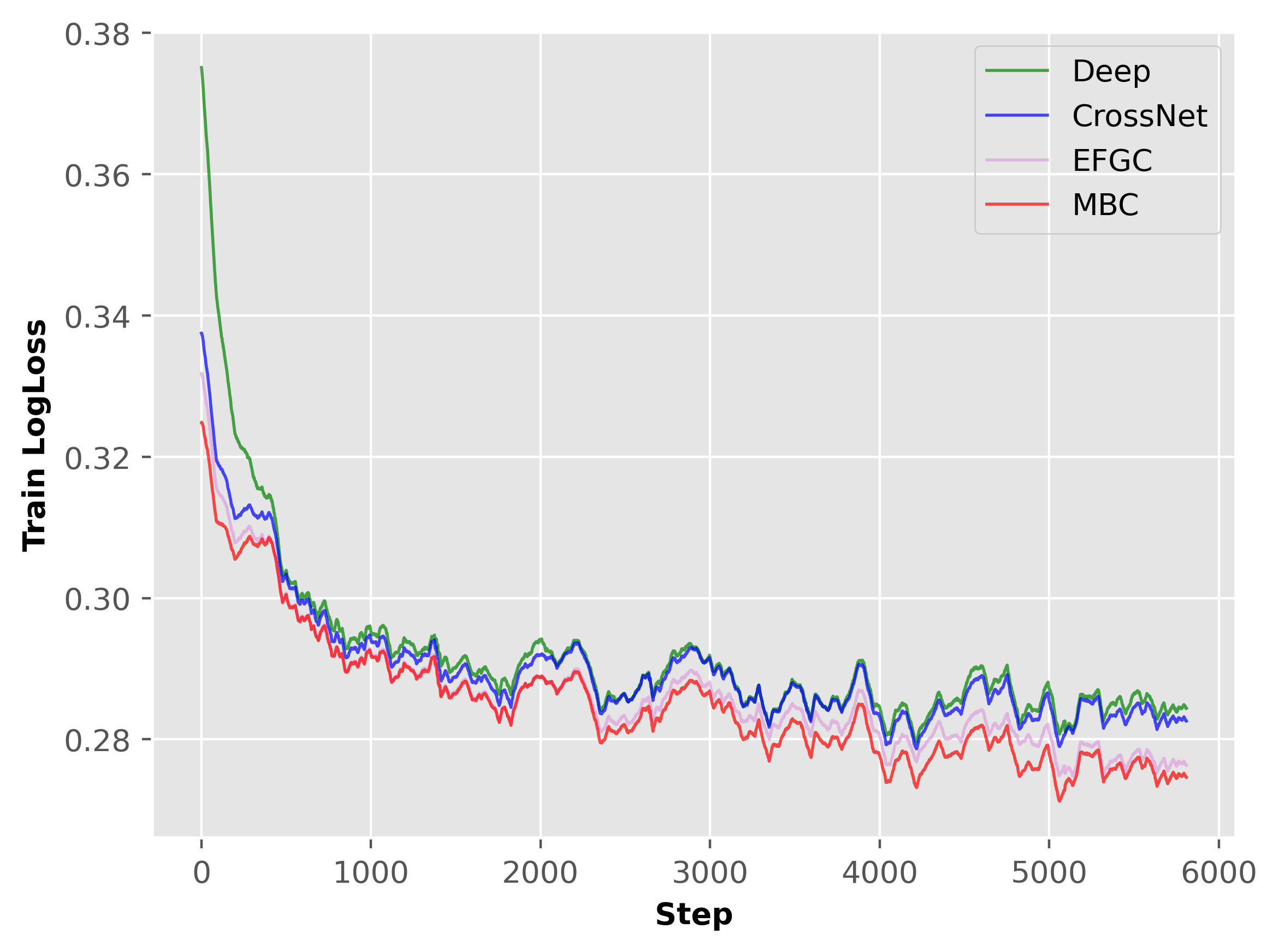}
\vspace{-20pt}
\caption*{\small{(a) Train LogLoss}}
\end{minipage}
\begin{minipage}[t]{0.22\textwidth}
\centering
\includegraphics[width=\textwidth]{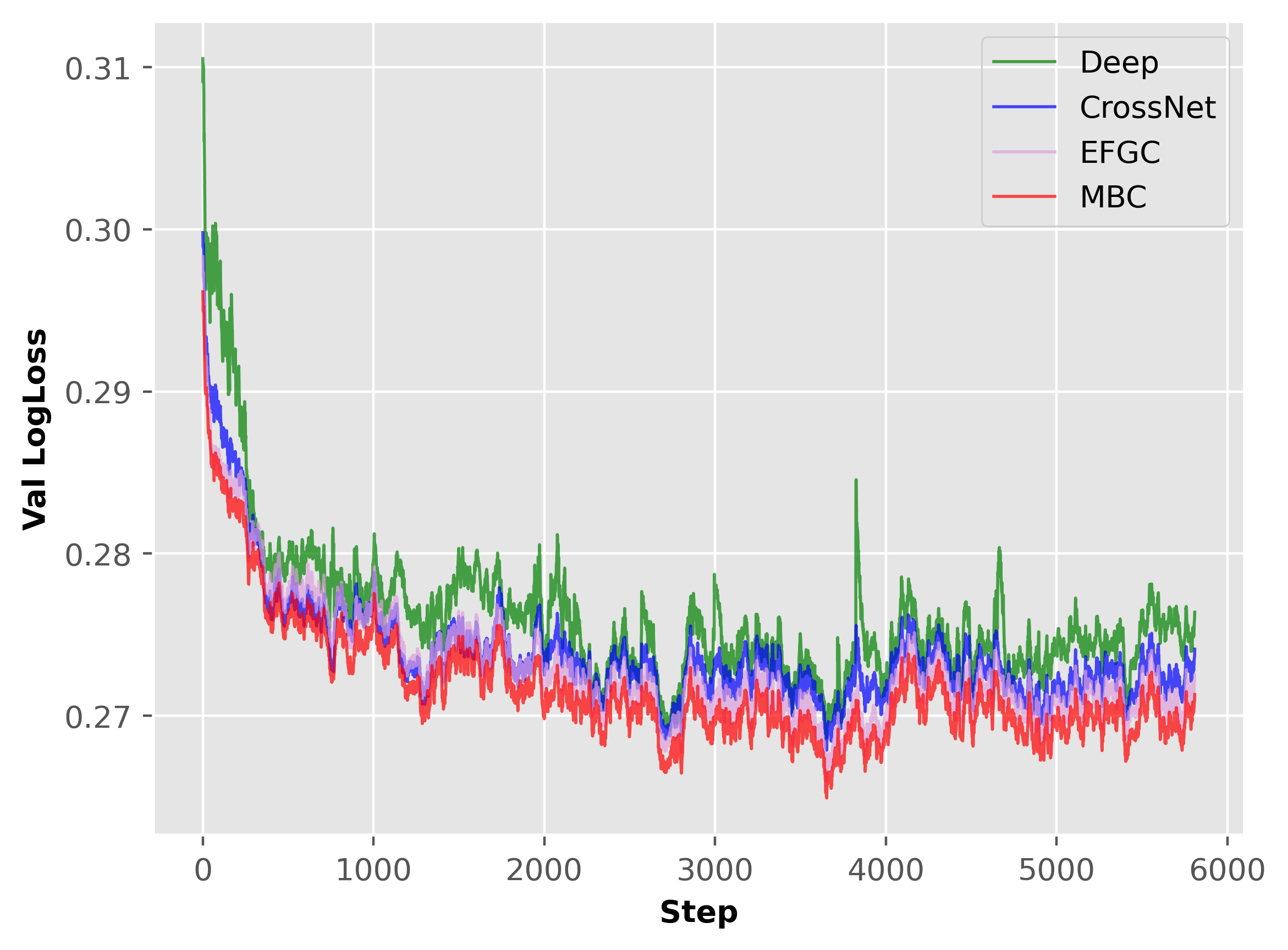}
\vspace{-20pt}
\caption*{\small{(b) Val LogLoss}}
\end{minipage}
\begin{minipage}[t]{0.22\textwidth}
\centering
\includegraphics[width=\textwidth]{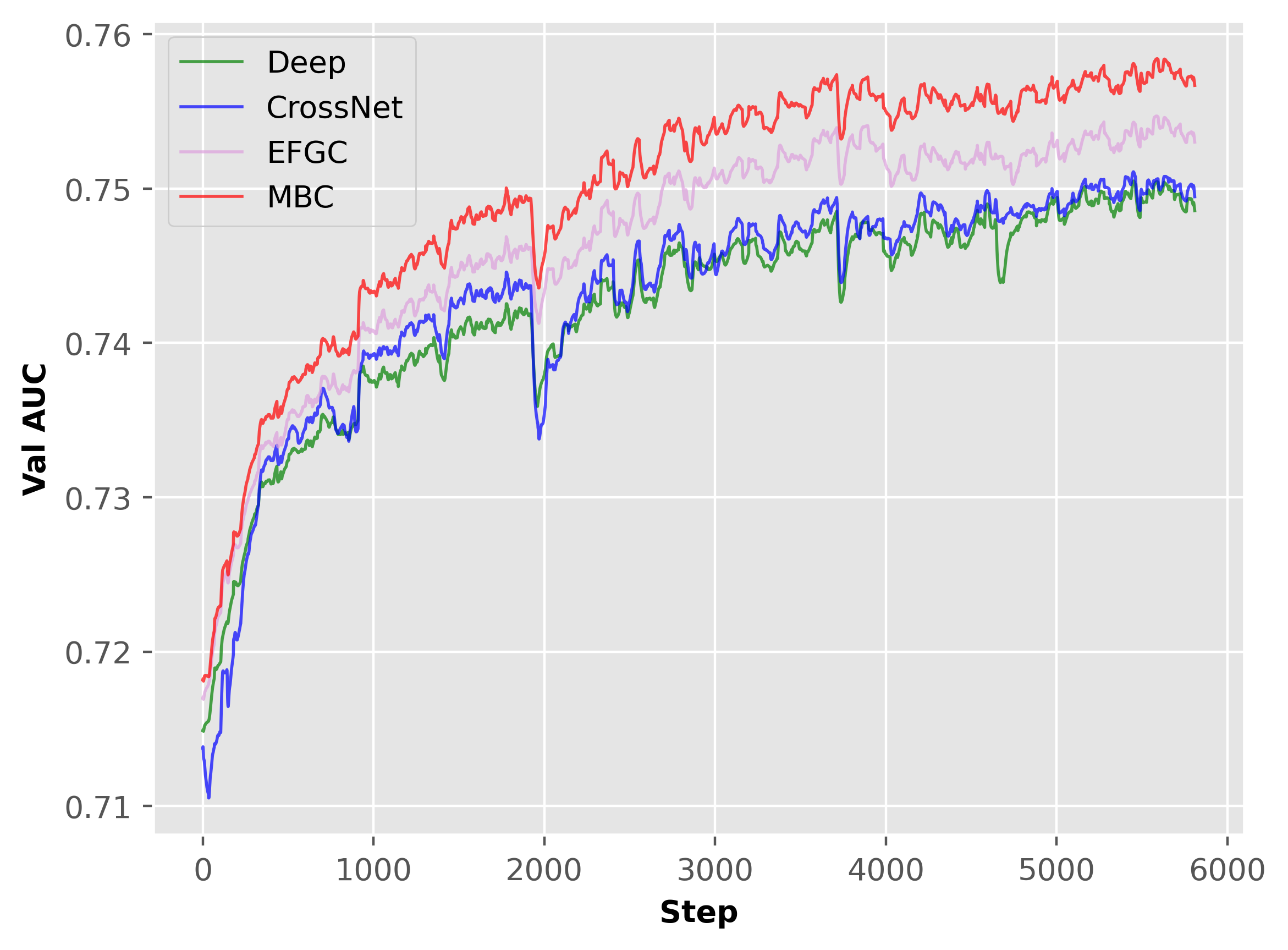}
\vspace{-20pt}
\caption*{\small{(c) Val AUC}}
\end{minipage}
\begin{minipage}[t]{0.22\textwidth}
\centering
\includegraphics[width=\textwidth]{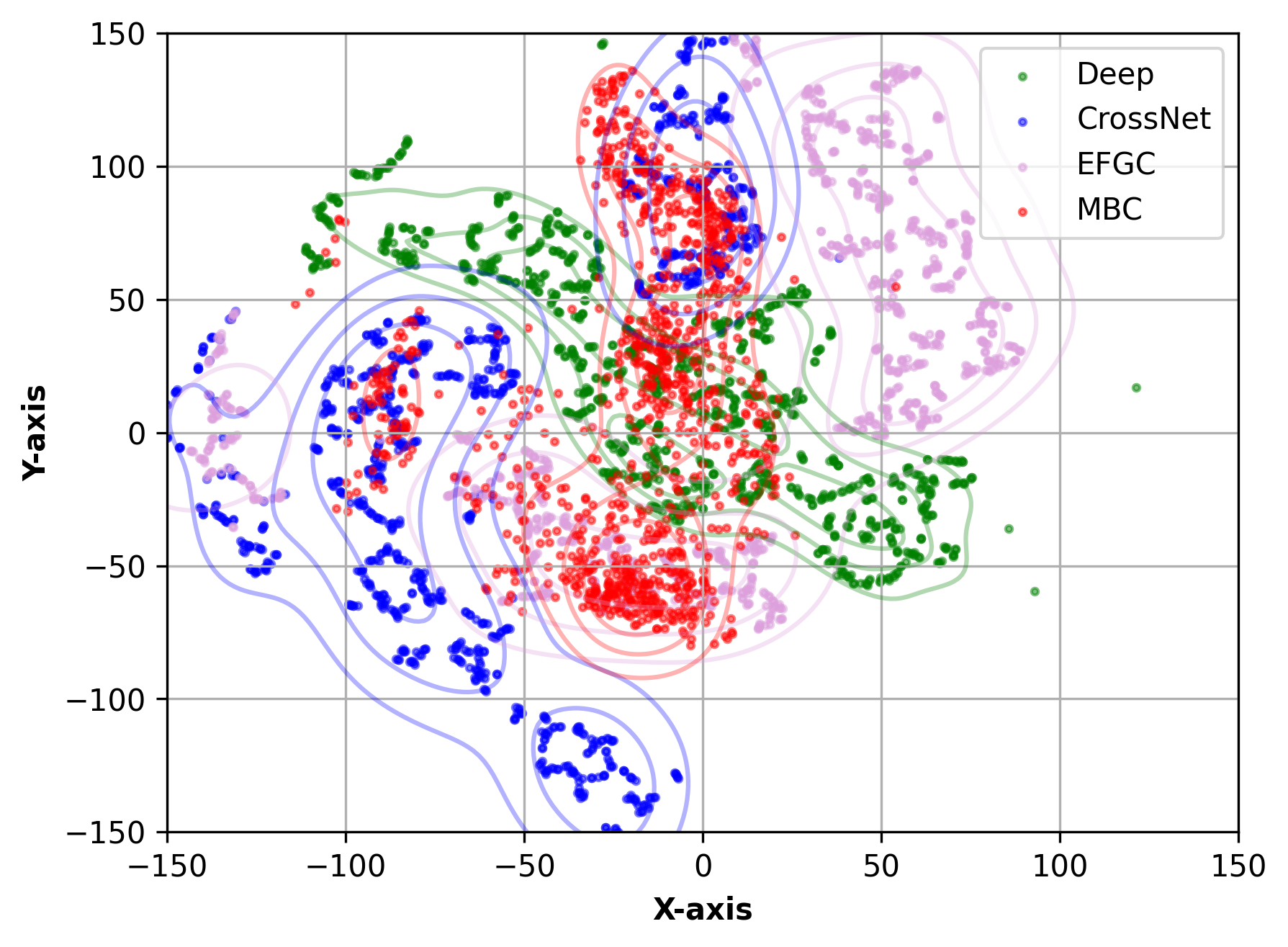}
\vspace{-20pt}
\caption*{\small{(d) Feature Distribution}}
\end{minipage}
\vspace{-8pt}
\caption{Comparison results of different branches on our Pailitao-12month dataset. (a)-(c) show the performance comparison of different branches in training. (d) indicates the out feature distribution of different branches after model convergence.}
\label{figure:why_mbc}
\vspace{-10pt}
\end{figure*}
\subsubsection{\textbf{Online Production Deployment}}
\label{sec:online_comparison}
We further evaluate the method through online A/B testing in image2product search of Alibaba Taobao system. In this experiment, the training data is our Pailitao-24month dataset. The baseline is our previously online serving model (DCNv2). Online evaluation metrics are real CTR, deal number by query, deal number by user and GMV. 
Table~\ref{table:product_results} shows that our MBCnet has an absolute \textbf{0.09 point} CTR increase, a relative \textbf{1.49\% }deal growth and a relative \textbf{1.62\%} GMV increase. 
In Figure~\ref{figure:online_AB}, we also present the CTR improvements compared to online base model across query categories and user groups. The figure clearly shows that MBCnet achieves significant improvements from multiple evaluation perspectives, highlighting its effectiveness in our production system. \textit{Since August 2024, MBCnet has been fully deployed online, serving hundreds of millions of customers.}

Meanwhile, different components of MBCnet work in a parallel style so that we can use parallel computation to accelerate training and inference. When conducting the online A/B test, the online base model is DCNv2 (\textit{i.e.} CrossNet and Deep branch), and MBCnet has additional EFGC branch for inference. \textbf{The EFGC branch is simple MLP and has a 2.2 million increased parameters, which is small for industrial CTR models.} \textbf{In our experiments, the real-time latency only has 1ms increase and the GPU utility only has 2\% increase, which are negligible in practice.} Detailed time complexity analysis is given in Appendix~\ref{sec:time_complexity}.

\subsection{Why Using Multi-Branch Cooperation}
\label{sec:why_multi_branch}
In this part, we conduct an experiment to demonstrate why using multi-branch cooperation in our industrial business. In particular, we show some learning dynamics of different branches during training. Moreover, we compare the feature distribution before logit layer across branches with t-SNE~\cite{tsne}. The results are summarized in Figure~\ref{figure:why_mbc}, where ``EFGC'', ``Deep'', ``CrossNet'' are three branches and ``MBC'' is the fused one.

From Figure~\ref{figure:why_mbc} (a)-(c), we see that: 1) 
The fused "MBC" model consistently achieves lower training log loss and validation log loss, along with a higher validation AUC throughout the training process. This indicates that a single branch has limited learning capacity, whereas MBCnet enhances its learning ability by cooperating multiple branches.
2) As shown in Figure~\ref{figure:why_mbc} (d), the output features from "EFGC," "Deep," and "CrossNet" branches are distinct from one another. In contrast, the fused "MBC" features are more evenly distributed across the feature space. In fact, we have also observed the output features across branches are more entangled and hard to be identified in early training stage, while the features tend to cluster to different subspaces and the fused ``MBC'' is becoming more evenly distributed across the feature space in later training epochs. \textbf{These cooperative dynamics demonstrate that each branch has its own modeling patterns, and MBCnet can gradually integrate their diverse capabilities.} 
These emphasize the value of utilizing collaboration among branches to model the complex distributions of user behaviors. \textbf{More experiments about why using multi-branch cooperation are given in Appendix~\ref{appendix:interaction_neurons} and~\ref{appendix:category_difference}.}

\begin{figure*}[t]
\centering
\begin{minipage}[t]{0.22\textwidth}
\centering
\includegraphics[width=\textwidth]{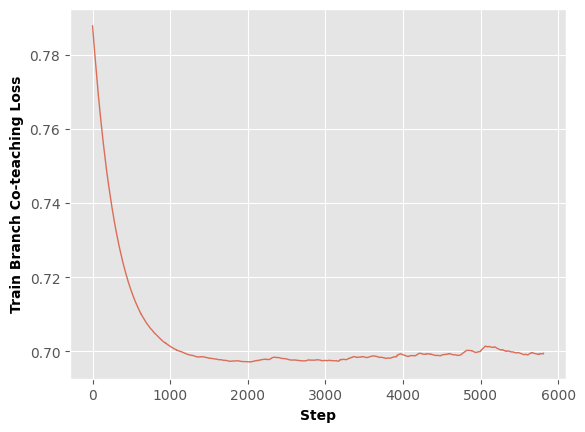}
\vspace{-20pt}
\caption*{{\small(a) Train Co-Teaching Loss}}
\end{minipage}
\begin{minipage}[t]{0.22\textwidth}
\centering
\includegraphics[width=\textwidth]{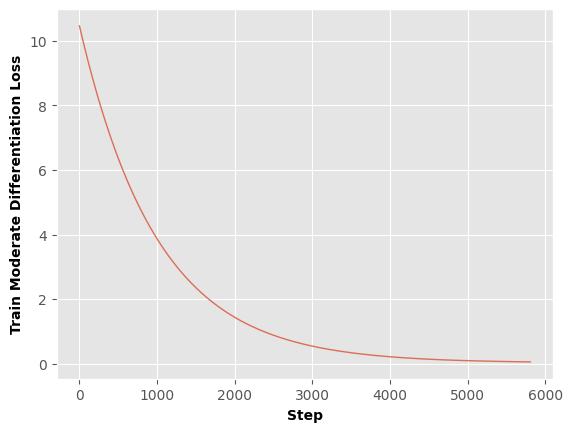}
\vspace{-20pt}
\caption*{\small{(b) Train Differentiation Loss}}
\end{minipage}
\begin{minipage}[t]{0.22\textwidth}
\centering
\includegraphics[width=\textwidth]{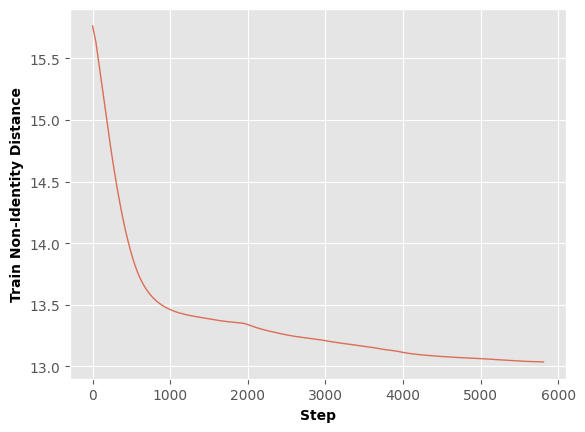}
\vspace{-20pt}
\caption*{\small{(c) Difference between $W$ and $I_{d}$}}
\end{minipage}
\begin{minipage}[t]{0.22\textwidth}
\centering
\includegraphics[width=\textwidth]{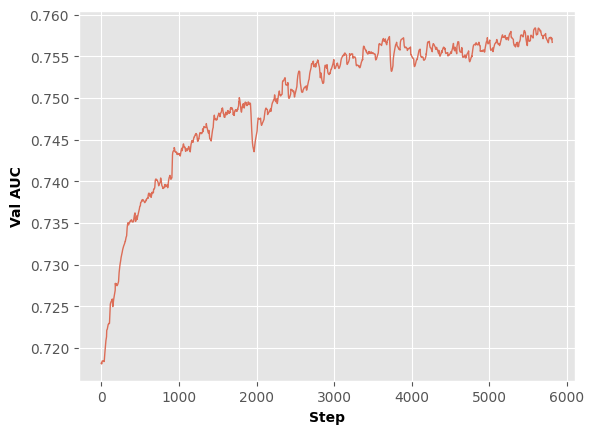}
\vspace{-20pt}
\caption*{\small{(d) Val AUC}}
\end{minipage}
\vspace{-8pt}
\caption{Different train and val metrics along training on our Pailitao-12month dataset. (a) and (b) show the train branch co-teaching loss $\mathcal{L}_{BCT}$ and train moderate differentiation loss $\mathcal{L}_{MDR}$, respectively. (c) indicates the Euclidean distance between equivalent transformation matrix $W$ and the identity matrix $I_{d}$. (d) shows the val auc of our MBCnet in different training steps.}
\label{figure:how_mbc}
\vspace{-10pt}
\end{figure*}

\begin{table}[]
\centering
\caption{Study of the cooperation principles to demonstrate how to do multi-branch cooperation.} 
\vspace{-8pt}
\label{table:study_principle}
\renewcommand{\arraystretch}{0.95}
 \setlength{\tabcolsep}{0.65mm}{ 
  \scalebox{0.8}{
\begin{tabular}{c|c|cc|cc}
\hline
\multirow{2}{*}{Study Principle} & Dataset                         & \multicolumn{2}{c|}{Pailitao-12month} & \multicolumn{2}{c}{Pailitao-24month} \\ \cline{2-6} 
                                 & Model Variant                   & AUC               & LogLoss           & AUC               & LogLoss          \\ \hline
\multirow{3}{*}{Principle 1}     & no discrimination               & 0.5772            & 0.3251            & 0.5575            & 0.3319           \\
                                 & weak to strong                  & 0.4933            & 0.3254            & 0.4754            & 0.3617           \\ \cline{2-6} 
                                 & \textbf{strong to weak}         & \textbf{0.7522}   & \textbf{0.2866}   & \textbf{0.7642}   & \textbf{0.2800}  \\ \hline
\multirow{3}{*}{Principle 2}     & max difference                  & 0.7176            & 0.3498            & 0.7366            & 0.3360           \\
                                 & min difference                  & 0.7285            & 0.3447            & 0.7349            & 0.3138           \\ \cline{2-6} 
                                 & \textbf{\textit{moderate differentiation}} & \textbf{0.7522}   & \textbf{0.2866}   & \textbf{0.7642}   & \textbf{0.2800}  \\ \hline
\end{tabular}
}}
\vspace{-15pt}
\end{table}
\subsection{How to do Multi-Branch Cooperation}
\label{sec:how_multi_branch}
We also explore various approaches in multi-branch cooperation to validate the effectiveness of our proposed two principles: \textbf{\textit{branch co-teaching}} and \textbf{\textit{moderate differentiation}}.
To examine the first principle, we modify our ``strong to weak'' learning scheme described in Eq.~\ref{eq:p_j_strong}-\ref{eq:co_teaching} to create two variants. The first variant, labeled ``no discrimination'', does not perform sample selection during co-teaching; instead, each branch can teach the other using the entire dataset. The second variant, labeled ``weak to strong'', reverses the teaching direction by having the weak branch instruct the strong branch. To investigate the second principle, we adjust the formulation of ``moderate differentiation'' from Eq.~\ref{eq:moderate_diff} in two ways. The first one, called ``max difference'', aims to maximize the L2 distance between latent features of branches. The second one, referred to as ``min difference'', seeks to minimize the L2 distance between branch features. The results are summarized in Table~\ref{table:study_principle}. We also show some key learning metrics during training to show the convergence and rationality of our MBCnet in Figure~\ref{figure:how_mbc}.

In Table~\ref{table:study_principle}, we show that violating our proposed scheme leads to a decline in model performance. Principle 1 is particularly crucial because it dictates the selection of knowledge to be transferred among branches. Principle 2 also affects performance by ensuring branch diversity and encouraging the model to discover different patterns. Additionally, Figure~\ref{figure:how_mbc} reveals that both the co-teaching loss and the moderate differentiation loss converge, while the validation AUC steadily increases. In the converged state shown in Figure~\ref{figure:how_mbc} (c), the difference between equivalent transformation matrix $W$ and $I_{d}$ remains non-zero, demonstrating that the branches maintain distinct discrepancies to explore diverse patterns.

\begin{table}[]
\centering
\caption{Ablation study of different model components.} 
\vspace{-8pt}
\label{table:ablation_study}
\renewcommand{\arraystretch}{0.9}
 \setlength{\tabcolsep}{1.0mm}{ 
  \scalebox{1.0}{
\begin{tabular}{c|cc|cc}
\hline
Dataset             & \multicolumn{2}{c|}{Pailitao-12month} & \multicolumn{2}{c}{Pailitao-24month} \\ \hline
Model               & AUC               & LogLoss           & AUC               & LogLoss          \\ \hline
w/o EFGC & 0.7445            & 0.2888            & 0.7548            & 0.2839           \\
w/o CrossNet & 0.7480            & 0.2880            & 0.7569            & 0.2828           \\
w/o DeepNet & 0.7501            & 0.2875            & 0.7622            & 0.2811           \\
w/o $\mathcal{L}_{\text{BCT}}$          & 0.7510            & 0.2874            & 0.7628            & 0.2806           \\
w/o $\mathcal{L}_{\text{MDR}}$          & 0.7506            & 0.2879            & 0.7603            & 0.2822           \\
w/o $\mathcal{L}_{\text{BCT}}$, w/o $\mathcal{L}_{\text{MDR}}$  & 0.7443            & 0.2954            & 0.7609            & 0.2818           \\ \hline
MBCnet              & \textbf{0.7522}   & \textbf{0.2866}   & \textbf{0.7642}   & \textbf{0.2800}  \\ \hline
\end{tabular}
}}
\vspace{-15pt}
\end{table}
\subsection{Study of Different Model Parts}
\label{sec:study_different_parts}
To evaluate the impact of each model component, we performed an ablation study by systematically removing specific modules, as detailed in Table~\ref{table:ablation_study}. \textit{The results indicate a clear decline in performance on both datasets when any component is removed.} For instance, excluding the EFGC branch leads to a 0.94\% decrease in AUC on Pailitao-24month. This result highlights the importance of our EFGC branch. The mechanism in EFGC highlights feature crossing within feature groups, which encourages the model to capture specific patterns and meanwhile reduces redundant crossing features. Additionally, removing the cooperation scheme, \textit{i.e.}, ``w/o $\mathcal{L}_{\text{BCT}}$, w/o $\mathcal{L}_{\text{MDR}}$'', results in a 0.79\% reduction in AUC on Pailitao-12month. 
These outcomes demonstrate that each component positively contributes to the overall performance of our model. \textbf{More experiments and analysis are provided in the Appendix.}

\section{Conclusion and Future Work}
In this paper, we introduced Multi-Branch Cooperation Network (MBCnet) for enhanced CTR prediction. MBCnet effectively integrates three distinct branches (\textit{i.e.}, EFGC, low rank CrossNet and Deep Net) in capturing complex feature interactions. \textit{The proposed cooperation scheme, based on the principles of \textbf{branch co-teaching} and \textbf{moderate differentiation}, facilitates meaningful collaboration between branches. It also enables the model to explore diverse feature interaction patterns and improve the overall prediction accuracy.} Our extensive experiments, including an online A/B test, demonstrated significant improvements in CTR, deal growth, and GMV, validating its effectiveness and scalability in real-world applications. 

Currently we simply use the common average pooling and use later MLP layer to implicitly combine the outputs of branches. We believe that a better way to fuse the branch outputs could lead to better performance. Meanwhile, we use loss values to classify branches as strong or weak on particular samples. This may produce unreliable loss measurements during the early training stage, potentially limiting the model's learning ability. We will study these issues in future work.


\bibliographystyle{ACM-Reference-Format}
\bibliography{sample-base}

\newpage
\appendix
\section{About Low-Rank Cross Net}
\label{appendix:dcnv2}
The low-rank cross network is from DCNv2~\cite{wang2021dcn}, and is a low-rank and computation-efficient version of the original cross network in DCN~\cite{10.1145/3124749.3124754}. The original cross network explicitly learn high-order feature interactions through defined equation $x_{l+1}=x_{0}x_{l}^{T}w_{l}+b_{l}+x_{l}$, where $x_{l},x_{l+1}\in \mathbb{R}^d$ are column vectors denoting the outputs from $l$-th and $(l+1)$-th cross layers, respectively. $w_{l},b_{l}\in \mathbb{R}^d$ are the weight and bias parameters of the $l$-th layer. The special structure of the cross network causes the degree of cross features to grow with layer depth. The highest polynomial degree (in terms of input $x_{0}$ for an $l$ layer cross network is $l+1$. In fact, the cross network comprises all the cross terms. Refer to~\cite{wang2021dcn} for detailed proof. 

This original cross network involves a dimension-consistent feature crossing in feature space, which would cause large computation when the dimension of inputs is high. Therefore, the low-rank cross network~\cite{wang2021dcn} aims to accelerate the feature interaction in original cross network. Specifically, it maps the feature into low-rank space and then maps it back to the original feature dimension with mixture of experts.

\begin{figure}[t]
\centering
\begin{minipage}[t]{0.22\textwidth}
\centering
\includegraphics[width=\textwidth]{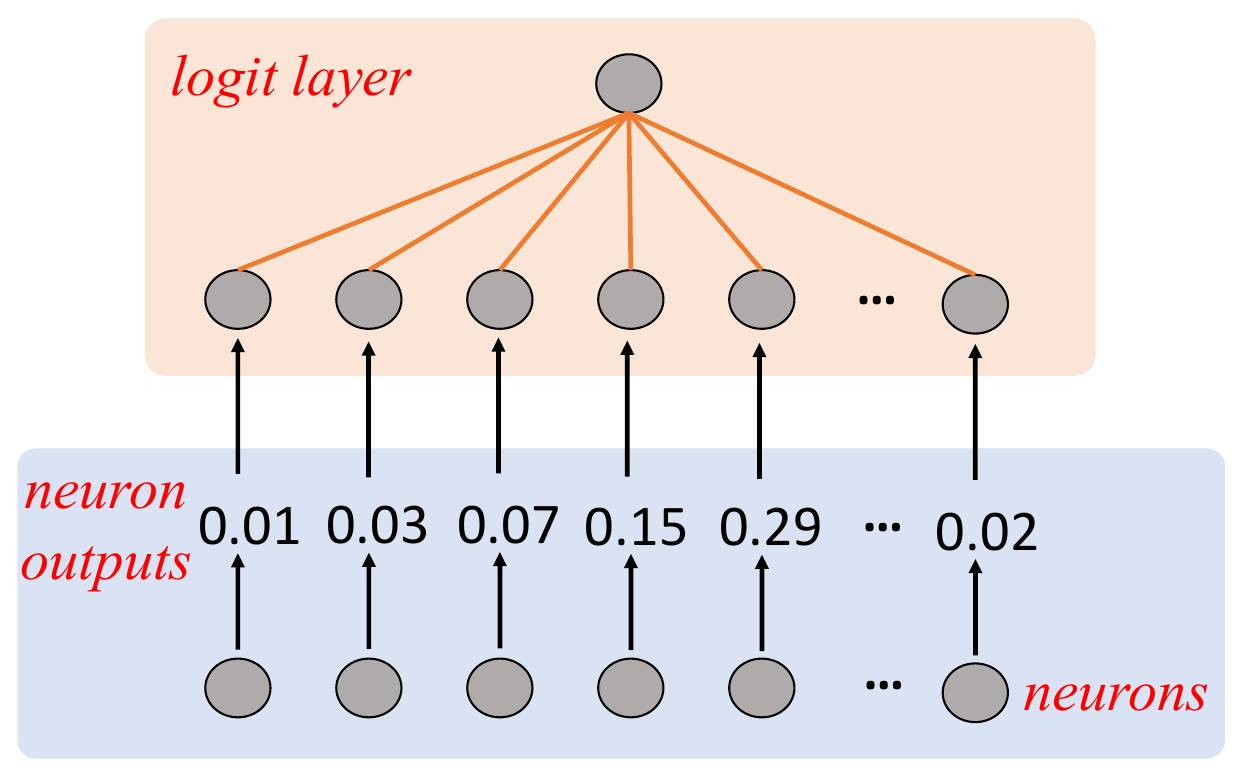}
\vspace{-20pt}
\caption*{(a)}
\end{minipage}
\begin{minipage}[t]{0.22\textwidth}
\centering
\includegraphics[width=\textwidth]{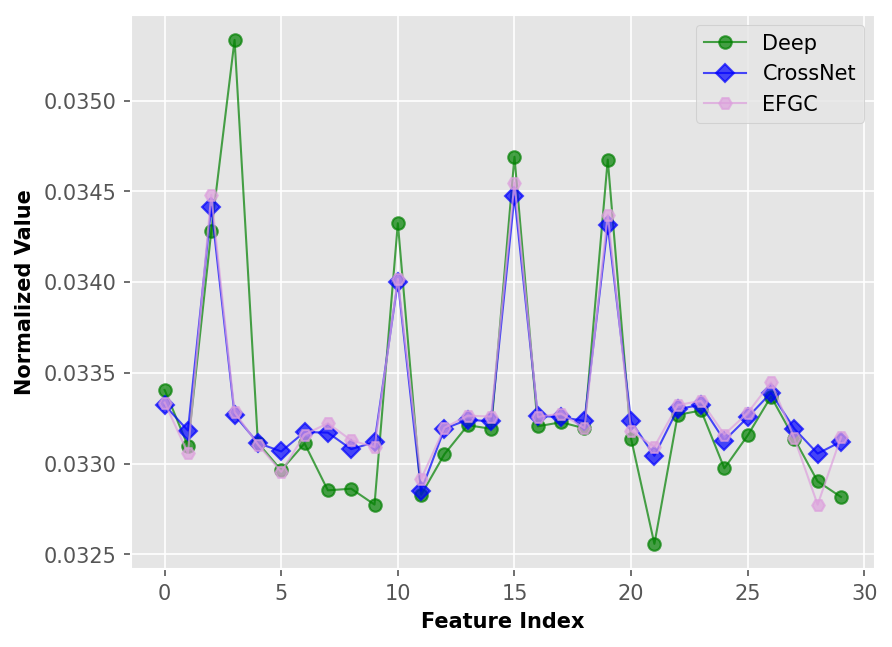}
\vspace{-20pt}
\caption*{(b)}
\end{minipage}
\vspace{-10pt}
\caption{(a) is an example to illustrate the structure of neuron outputs before logit layer. (b) shows the output distribution of those neurons on Pailitao-12month.}
\label{figure:softmax_outputs_of_neurs}
\vspace{-10pt}
\end{figure}
\section{Additional Experiments}
\subsection{Branch Difference at Neuron Level}
\label{appendix:interaction_neurons}
We design an experiment to illustrate the branch difference at neuron level. In particular, MBCnet integrates latent features from multiple branches to execute the final CTR prediction task. These latent features are derived from the outputs of neurons, as depicted in Figure~\ref{figure:softmax_outputs_of_neurs} (a). Each dimension of a feature value reflects the strength of a specific high-order feature.
To analyze these feature strengths for each branch, we normalize the neuron outputs using softmax function\footnote{Our network has an output dimension of 128. For clarity, we randomly select 30 neuron outputs to illustrate the results.}. The normalized distributions are shown in Figure~\ref{figure:softmax_outputs_of_neurs} (b)\footnote{Notice that the output features from different branches are in the same space in our network design, enabling direct comparison of neuron values across branches.}.

From the figure, we have the following observations: 1) Different branches have close outputs at most neurons while quite different values at few neurons. This indicates different branches have many common learned patterns and several specific patterns, which aligns with our moderate differentiation principle. \textit{The common patterns ensure branches are learned towards the same learning objective, while the specific patterns encourage the model to develop complementary functionalities, benefiting the final prediction task.} 2) We also observe that ``EFGC'' branch exhibits greater consistency with ``CrossNet'' branch compared to ``Deep'' branch. One possible reason for this may be that the ``EFGC'' branch engages in feature crossing by domain-driven knowledge, which can be regarded as one kind of explicit feature crossing in ``CrossNet''.

\begin{figure}[t]
\centering
\begin{minipage}[t]{0.32\textwidth}
\centering
\includegraphics[width=\textwidth]{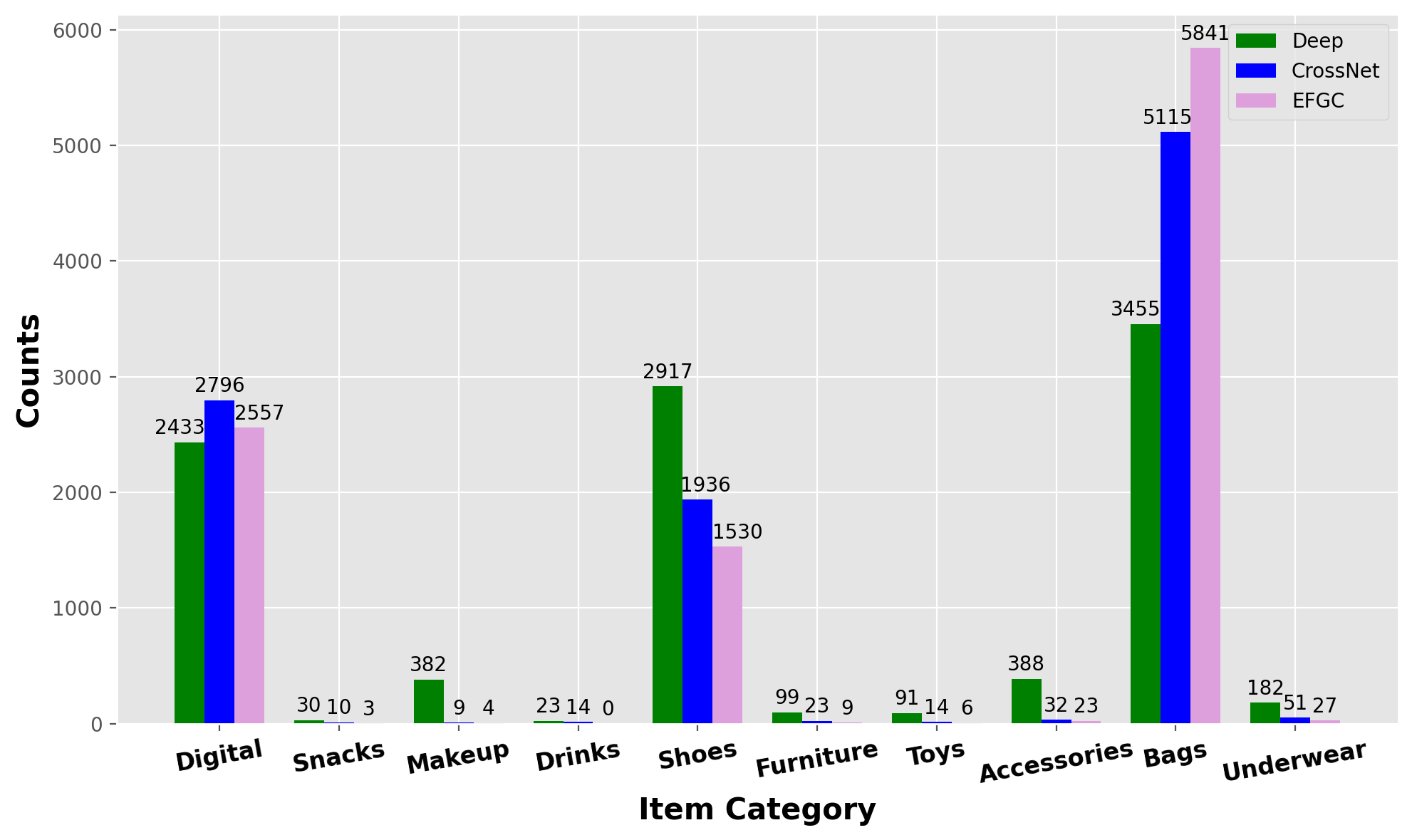}
\vspace{-20pt}
\end{minipage}
\vspace{-10pt}
\caption{The category distribution of top-10k most accurately learned samples (\textit{i.e.}, those with low log loss) across different branches in our MBCnet.}
\label{figure:category_learning_difference}
\vspace{-10pt}
\end{figure}
\subsection{Branch Difference at Category Level}
\label{appendix:category_difference}
In this part, we conduct an experiment to highlight the branch learning difference at category level. Specifically, for each branch, we first sort the samples in ascending order based on their logloss values, where a lower log loss value indicates an accurately learned sample. We then select the top 10,000 samples and count them under each category to get the category distribution of these accurately learned samples for each branch. The results are presented in Figure~\ref{figure:category_learning_difference}.

From this figure, we clearly see that \textit{different branches have unique learning strengths and prefer distinct categories.} For example, “Deep” branch outperforms both ``CrossNet'' and ``EFGC'' branches in learning ``Makeup'', ``Shoes'', ``Accessories'' and ``Underwear'' categories, but it exhibits weaker performance in ``Digital'' and ``Bags'' categories. In contrast, the proposed ``EFGC'' branch excels over the other two branches in ``Bags'' category but falls behind in ``Shoes'' and ``Snacks'' categories. \textit{These findings highlight the unique advantages of each branch and their complementary roles across different types of data.}

\begin{figure}[t]
\centering
\begin{minipage}[t]{0.3\textwidth}
\centering
\includegraphics[width=\textwidth]{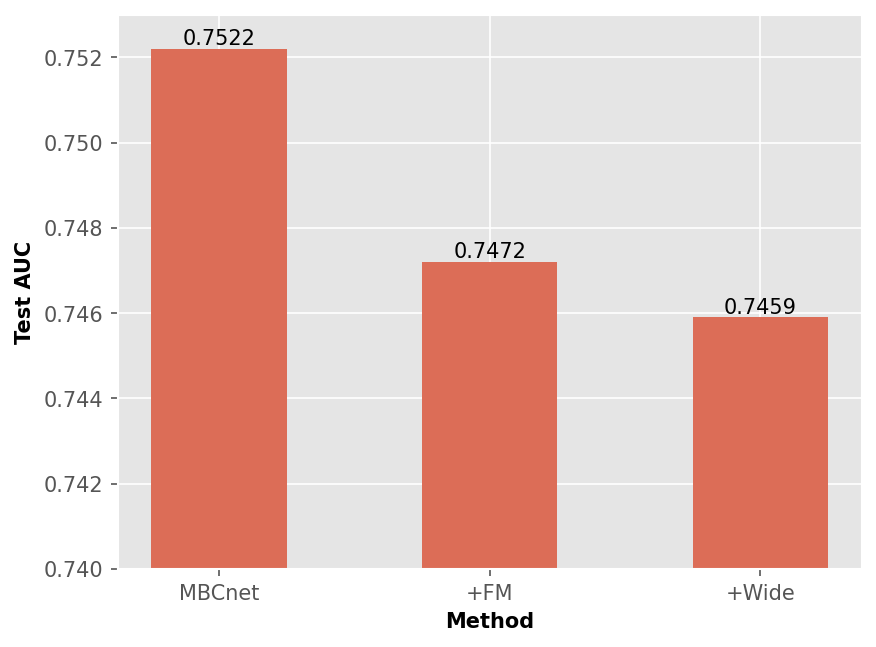}
\vspace{-20pt}
\end{minipage}
\vspace{-8pt}
\caption{Model performance when adding more branches on Pailitao-12month.}
\label{figure:add_fm_wide}
\vspace{-10pt}
\end{figure}
\subsection{Adding More Branches into MBCnet}
In this part, we further add two popular feature interaction branches, \textit{i.e.}, factorization machine in DeepFM~\cite{10.5555/3172077.3172127} and wide network in Wide\&Deep~\cite{cheng2016wide}, and observe the model performance. The results are summarized in Figure~\ref{figure:add_fm_wide}. This figure indicates that incorporating FM and Wide results in a decline in performance. The primary reason for this is the complexity of the distribution of our extensive user behavior data, which means that these lower-order techniques fail to deliver any benefits compared to the other branches.

\begin{figure}[t]
\centering
\begin{minipage}[t]{0.22\textwidth}
\centering
\includegraphics[width=\textwidth]{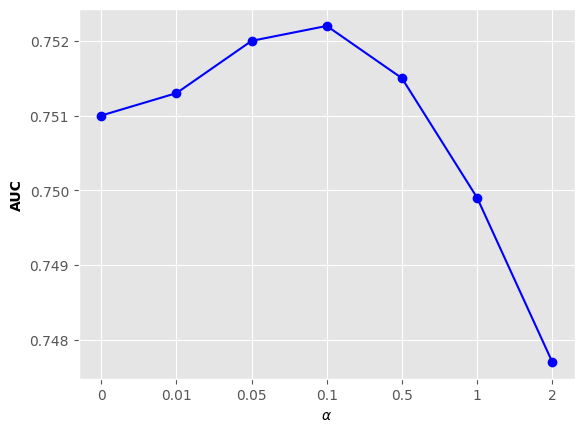}
\vspace{-20pt}
\caption*{(a)}
\end{minipage}
\begin{minipage}[t]{0.22\textwidth}
\centering
\includegraphics[width=\textwidth]{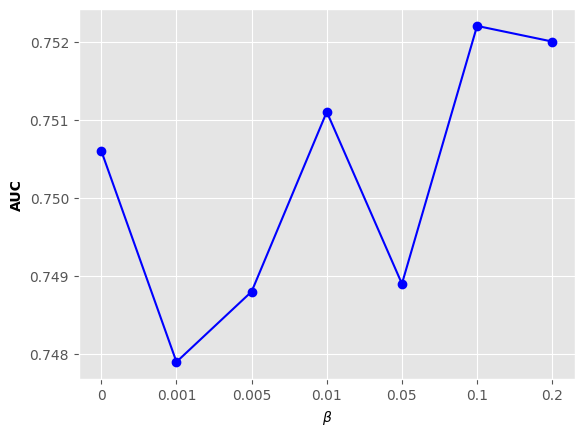}
\vspace{-20pt}
\caption*{(b)}
\end{minipage} 
\vspace{-10pt}
\caption{Hyper-parameter sensitivity on Pailitao-12month.}
\label{figure:hyperparameter}
\vspace{-10pt}
\end{figure}
\subsection{Hyper-Parameter Sensitivity}
In MBCnet, the hyper-parameters $\alpha$ and $\beta$ regulate the weighting of the loss functions $\mathcal{L}_{\text{BCT}}$ and $\mathcal{L}_{\text{MDR}}$, respectively. To understand the impact of these parameters, we performed a sensitivity analysis, and the results are illustrated in Figure~\ref{figure:hyperparameter}. 
From the figure, we observe that: 1) Comparing the results in Figure~\ref{figure:hyperparameter} with those in Table~\ref{table:offline_overall_comparison}, it is clear that our model can perform better than the baselines across a wide range of $\alpha$ and $\beta$ values. This robustness highlights the effectiveness of incorporating cooperative interactions between different branches of the model. 2) The performance of the model degrades if $\alpha$ and $\beta$ are set either too high or too low. Specifically, excessively large values for these hyper-parameters can dominate the primary CTR task, while too small values may make the cooperative effects between branches negligible. 

\begin{table}[]
\caption{The AUC comparison of different models under different data density on Pailitao-12month. The listed density ratio is relative to the original train set.}
\label{tab:different_sparsity}
\centering
\begin{tabular}{l|lllll}
\hline
Density ratio   & 20\%            & 40\%            & 60\%            & 80\%            & 100\%           \\ \hline
DNN             & 0.5132          & 0.5885          & 0.6770          & 0.7223          & 0.7423          \\ \hline
DCNv2           & 0.5200          & 0.5929          & 0.6797          & 0.7262          & 0.7461          \\ \hline
\textbf{MBCnet} & \textbf{0.5277} & \textbf{0.6015} & \textbf{0.6890} & \textbf{0.7319} & \textbf{0.7522} \\ \hline
\end{tabular}
\end{table}
\subsection{Performance Under Different Data Sparsity}
We also conducted performance comparison under different data density. Specifically, we keep the validation and test set fixed, and mask a certain ratio of positive samples. The AUC results of some methods on Pailitao-12month are given in Table~\ref{tab:different_sparsity}. The result shows that MBCnet consistently achieves better performance even in more sparse cases.

\section{Time Complexity Analysis}
\label{sec:time_complexity}
In Table~\ref{tab:complexity}, we have made a complexity analysis of different branches/components in MBCnet. The time complexity of EFGC branch is $\mathcal{O}(N_{g}L_{\text{EFGC}}d^{2})$, where $N_{g}$ is the number of feature groups, $L_{\text{EFGC}}$ is the number of EFGC layers and $d$ denotes the latent dimension. The time complexity of CrossNet branch is $\mathcal{O}(L_{\text{CrossNet}}N_{\text{experts}}Fd)$, where $L_{\text{CrossNet}}$ is the number of CrossNet layers, $N_{\text{experts}}$ denotes the number of experts and $F$ is the input feature dimension. For Deep branch, the time complexity is $\mathcal{O}(L_{\text{Deep}}d^2)$, where $L_{\text{Deep}}$ is the number of mlp layers. Considering the branch fusion in Eq.15 is average pooling and the complexity is negligible, so the complexity of MBCnet can be taken as the summation of three branches. 

\textbf{Fortunately, different components of MBCnet works in a parallel style so that we can use parallel computation to accelerate training and inference.} \textbf{In our experiments, the real-time latency only has 1ms increase and the GPU utility only has 2\% increase, which are negligible in practice.}

\begin{table}[]
\centering
\caption{Time complexity analysis of different model components.} 
\vspace{-8pt}
\label{tab:complexity}
\renewcommand{\arraystretch}{1.5}
 \setlength{\tabcolsep}{0.7mm}{ 
  \scalebox{0.6}{
\begin{tabular}{c|ccc}
\hline
Method      & EFGC                                                                                                                                                      & CrossNet                                                                                                                                                                       & Deep                                        \\ \hline
Complexity  & $\mathcal{O}(N_{g}L_{\text{EFGC}}d^{2})$                                                                                                                  & $\mathcal{O}(L_{\text{CrossNet}}N_{\text{experts}}Fd)$                                                                                                                         & $\mathcal{O}(L_{\text{Deep}}d^2)$           \\ \hline
Ilustration & \begin{tabular}[c]{@{}c@{}}$N_{g}$: number of feature groups\\ $L_{\text{EFGC}}$: number of EFGC layers\\ $d$: latent dimension.\end{tabular} & \begin{tabular}[c]{@{}c@{}}$L_{\text{CrossNet}}$: number of CrossNet layers\\ $N_{\text{experts}}$: number of experts\\ $F$: input feature dimension.\end{tabular} & $L_{\text{Deep}}$: number of mlp layers \\ \hline
\end{tabular}
}}
\vspace{-15pt}
\end{table}

\section{Core Codes of MBCnet}
\label{appendix:core_code}
To better illustrate the scheme of our cooperation idea, we provide the Tensorflow codes in Listing~\ref{lst:mbc_code}. The loss\_op function defines the entrance of our training objective. 
\onecolumn

\definecolor{codeblue}{rgb}{0.25,0.5,0.5}
\definecolor{codekw}{rgb}{0.85, 0.18, 0.50}
\lstset{
  backgroundcolor=\color{white},
  basicstyle=\fontsize{9pt}{9pt}\ttfamily\selectfont,
  columns=fullflexible,
  breaklines=true,
  caption={~Loss Code of MBCnet in Tensorflow Style},
  captionpos=t,
  commentstyle=\fontsize{9pt}{9pt}\color{codeblue},
  keywordstyle=\fontsize{9pt}{9pt}\color{codekw},
}
\begin{lstlisting}[language=python,label={lst:mbc_code}]
import tensorflow as tf

# binary cross entropy loss
def bce_loss_op(self, logit, label, reduction='mean'):
    if reduction=='mean':
        bce_loss = tf.reduce_mean(tf.nn.sigmoid_cross_entropy_with_logits(logits=logit,labels=label))
    elif reduction=='sum':
        bce_loss = tf.reduce_sum(tf.nn.sigmoid_cross_entropy_with_logits(logits=logit,labels=label))
    elif reduction=='none':
        bce_loss = tf.nn.sigmoid_cross_entropy_with_logits(logits=logit,labels=label)
    else:
        print('Not supported reduction!')
        raise Exception
    return bce_loss

# branch co-teaching loss
def BCT_loss_op(self, logits_list, labels):
    losses_list = []
    for logits in logits_list:
        loss = self.bce_loss_op(logits, labels, reduction='none')
        losses_list.append(loss)
    BCT_loss = 0.0
    n_sample = 1e-8
    for i in range(len(logits_list)):
        for j in range(i,len(logits_list)):
            if i!=j:
                # mask of disagreement data
                ###### simplified version
                mask_i = tf.logical_and(tf.reshape(tf.cast(losses_list[i] < -tf.log(0.5), tf.bool), [-1]), tf.reshape(tf.cast(losses_list[j] > -tf.log(0.5), tf.bool),[-1]))
                mask_j = tf.logical_and(tf.reshape(tf.cast(losses_list[j] < -tf.log(0.5), tf.bool), [-1]),tf.reshape(tf.cast(losses_list[i] > -tf.log(0.5), tf.bool),[-1]))

                # label supervision from strong branch
                BCT_loss += tf.cond(tf.reduce_sum(tf.cast(mask_j,tf.float32))>0, lambda:self.bce_loss_op(tf.boolean_mask(logits_list[i],mask_j),tf.stop_gradient(tf.boolean_mask(tf.sigmoid(logits_list[j]),mask_j)),reduction='sum'), lambda:tf.zeros_like(self.reg_loss))
                BCT_loss += tf.cond(tf.reduce_sum(tf.cast(mask_i,tf.float32))>0, lambda:self.bce_loss_op(tf.boolean_mask(logits_list[j],mask_i),tf.stop_gradient(tf.boolean_mask(tf.sigmoid(logits_list[i]),mask_i)),reduction='sum'), lambda:tf.zeros_like(self.reg_loss))
                n_sample += tf.reduce_sum(tf.cast(mask_j,tf.float32)) + tf.reduce_sum(tf.cast(mask_i,tf.float32))
    BCT_loss = BCT_loss/n_sample
    return BCT_loss

# moderate differentiation loss
def MDR_loss_op(self, feats_list):
    mdr_loss_list = []
    hidden_dim = feats_list[0].get_shape().as_list()[1]
    for i in range(len(feats)):
        for j in range(len(feats)):
            if i!=j:
                feat_i, feat_j = feats[i], feats[j]
                # orthogonal transformation constraint
                # W_{ji} is the transpose of W_{ij}
                mapped_feat_i_to_j = tf.matmul(feat_i, self.params['W_{}{}'.format(i,j)])
                mapped_feat_i_to_j_to_i = tf.matmul(mapped_feat_i_to_j, self.params['W_{}{}'.format(j,i)])
                l2_distance = tf.norm(mapped_feat_i_to_j-feat_j, ord='euclidean', axis=1) + tf.norm(mapped_feat_i_to_j_to_i-feat_i, ord='euclidean', axis=1)
                mdr_loss_list.append(l2_distance)
    mdr_loss = tf.reduce_mean(tf.concat(mdr_loss_list, 0))
    return mdr_loss

# The overall loss op
def loss_op(self, branch_logits_list, branch_top_feats_list, fused_logits, labels):
    '''
    branch_logits_list: a list of branch output logits, [efgc_logits, crossnetv2_logits, dnn_logits], each one is [B, 1]
    branch_top_feats_list: a list of branch output features before logit layer, [efgc_top_feat, crossnetv2_top_feat, dnn_top_feat], each one is [B, 1]
    fused_logits: the fused branch logits, [B, 1]
    labels: the labels of samples, [B, 1]
    '''
    efgc_logits, crossnetv2_logits, dnn_logits = branch_logits_list    
    with tf.name_scope("CTR_Loss_Op"): 
      # EFGC bce loss
      efgc_bce_loss = self.bce_loss_op(efgc_logits)
      # CrossNetV2 bce loss
      crossnetv2_bce_loss = self.bce_loss_op(crossnetv2_logits)
      # DNN bce loss
      dnn_bce_loss = self.bce_loss_op(dnn_logits)
      # fused bce loss
      fused_bce_loss = self.bce_loss_op(fused_logits)
      # overall bce loss
      self.bce_loss = efgc_bce_loss + crossnetv2_bce_loss + dnn_bce_loss + fused_bce_loss

      # multi-branch cooperation loss 
      self.BCT_loss = self.BCT_loss_op(branch_logits_list, labels)
      self.MDR_loss = self.MDR_loss_op(branch_top_feats_list)

      # final optimization loss
      self.loss = self.bce_loss + self.alpha * self.BCT_loss + self.beta * self.MDR_loss
\end{lstlisting}
\twocolumn

\end{document}